
\magnification=1200
\baselineskip=18.0truept
\ \medskip
\centerline{{\bf ANTISYMMETRIZATION OF A MEAN FIELD CALCULATION}}
\centerline{{\bf OF THE T-MATRIX}}
\medskip
\centerline{A.Wierling}
\centerline{Max-Planck Gesellschaft, AG "Theoretische
Vielteilchensysteme", 18051 Rostock, Germany}
\smallskip
\centerline{B.G.Giraud and $ {\rm F.Mekideche}^\ast $}
\centerline{Service Physique Th\'eorique, DSM-CEA Saclay,
91192 Gif/Yvette,
France}
\smallskip
\centerline{H.Horiuchi, T.Maruyama$^{\ast\ast}$ and A.Ohnishi$^{\ast\ast\ast}$}
\centerline{Physics Department, Kyoto University, Kyoto 606, Japan}
\smallskip
\centerline{and}
\centerline{J.C.Lemm, A.Weiguny}
\centerline{Institut f\"ur Theoretische Physik,
Universit\"at M\"unster, 48149
M\"unster, Germany}
\medskip
\noindent{\bf ABSTRACT} \par
The usual definition of the prior(post) interaction
$ V(V^\prime ) $ between
projectile and target (resp. ejectile and residual target) being
contradictory with full antisymmetrization
between nucleons, an explicit
antisymmetrization projector $ {\cal A} $ must be included
in the definition
of the
transition operator,
$ T\equiv V^\prime{\cal A}+V^\prime{\cal A}GV. $ We
derive the suitably antisymmetrized
mean field equations leading to a non perturbative estimate of $ T$.
The theory is illustrated by a
calculation of forward $\alpha$-$\alpha$
scattering,
making use of self consistent symmetries.


\vskip3.5cm

\hrule
\vskip0.3cm
\noindent
$\ast$  {\it Permanent address:} Institut de Physique, Universit\'e Sciences et
Technologie,
BP 32  El Alia, Alger, Alg\'erie

\noindent
$\ast\ast$ {\it Permanent address:} Advanced Science Research Center, Japan
Atomic Energy
Research Institute, Tokai, Ibaraki 319-11, Japan

\noindent
$\ast\ast\ast$ {\it Permanent address:} Department of Physics, Hokkaido
University, Sapporo
060, Japan

\smallskip
\vfill\eject
\smallskip

\centerline{1. Introduction}
\medskip
Consider the usual Hamiltonian $ H= \sum^{ }_ it_i+ \sum^{ }_{ i>j}v_{ij}, $
describing a
two-fragment collision a+b$\rightarrow$c+d between $ N $ nucleons, elastic or
inelastic,
with or without rearrangement. It is customary to define the prior (post)
interaction $ V= \sum^{ }_{ i\epsilon a,j\epsilon b}v_{ij}, $ $ (V^\prime =
\sum^{ }_{ i\epsilon c,j\epsilon d}v_{ij}), $ respectively, at the cost of
considering as distinct the nucleons of the projectile a (resp. ejectile c)
with respect to the target b (resp. residual target d) in the initial
(resp. final) channel. Under this lack of antisymmetrization, one defines
an initial wave $ \chi , $ which is a straight product of plane waves for a
and b.
Similarly one defines a non antisymmetrized final wave $ \chi ^\prime $ for c
and d. The
transition amplitude is then introduced as the on-shell matrix element
$ <\chi ^\prime \mid (V^\prime +V^\prime GV)\mid \chi >, $ where $ G $ is the
full $ N- {\rm nucleon} $ Green function at the
energy $ E $ of the collision. \par
\smallskip
In order to take the Pauli principle into account, the correct
approach, as described for instance by $ {\rm Austern} [1], $ consists in
{\sl i\/}) retaining
$ \chi ,\chi ^\prime $ as non antisymmetrized, but {\sl ii\/}) calculating the
matrix element
$ <\chi ^\prime \mid (V^\prime{\cal A}+V^\prime{\cal A}GV)\mid \chi >, $ where
the projector $ {\cal A}=(N!)^{-1} \sum^{ }_{{\cal P}}(-)^{{\cal P}}{\cal P},
$ namely the parity
weighted sum of all the permutations $ {\cal P} $ of $ N $ nucleons, takes
care of the
antisymmetrization.
Note that the validity of the above formula for the antisymmetrized
transition amplitude rests upon the assumption that initial and final
wave packets $\chi$, $\chi^\prime$ are normalized to unity and describe
non-overlapping fragments $a, b$ and $c, d$, respectively.
Asymptotic density and
flux are then unchanged under antisymmetrization. We shall come back to
this point in Section 5.
\par
\smallskip
The calculation of Born terms, antisymmetrized or not, being
straightforward, the present paper focusses on the following {\sl generic\/}
problem:
\noindent Let $ z $ be any complex energy (eventually $ z\rightarrow E+i0). $
Let $ \chi_ a({\vec r}_1...{\vec r}_{N_a}) $ be any
wave function for the projectile, with internal antisymmetrization for
these $ N_a $ nucleons. For instance, $ \chi_ a $ may be a boosted shell model
wave
function, with or without configuration mixing, but it can actually differ
arbitrarily from an eigenstate of the projectile Hamiltonian $ H_a. $ Let
$ \chi_ b({\vec r}_{N_a+1}...{\vec r}_N) $ be an arbitrary, antisymmetrized
wave
 function
for nucleus
b, and consider the straight product $ \chi =\chi_ a\chi_ b. $ For the final
state, consider
similarly $ \chi^{ \prime}_ c, $ $ \chi^{ \prime}_ d $ and $ \chi ^\prime
=\chi^{ \prime}_ c\chi^{ \prime}_ d, $ with antisymmetrization internal to c
and d,
separately. We want to calculate the number
\smallskip
$$ {\cal D}(z)=<\chi ^\prime \mid V^\prime{\cal A}(z-H)^{-1}V\mid \chi >.
\eqno (1.1) $$

\smallskip
The main motivation for this problem is, obviously, that $ {\cal D} $
represents
the multistep transition amplitude $ (T- {\rm matrix} $ without Born term)
when $ \chi ,\chi ^\prime $
are channel waves and $ z $ reaches the retarded on shell limit. But the
validity of the arguments listed in this paper extends to all complex
values of $ z $ and very general choices for $ \chi ,\chi ^\prime . $ (The
only restrictive
condition for this validity is the square integrability of $ V\mid \chi > $
and
$ V^\prime \mid \chi ^\prime >, $ so that $ {\cal D} $ be well defined as a
finite number. It is automatically
satisfied for two-fragment channels and short range potentials.) \par
\smallskip
A variational calculation of $ {\cal D} $ is described in Section 2. Then we
introduce in Section 3 a time independent mean field (TIMF) approximation [2]
for this calculation.
Section 4 is a discussion of symmetries which can simplify the variational
TIMF equations.
A numerical application of the theory to $\alpha$-$\alpha$
scattering is
described in Section 5, making use of self consistent symmetries introduced
in Section 4.
The final Section 6 contains our discussion and
conclusion.
Finally we introduce in an Appendix a new formalism for the representation
of nucleons in different clusters as fermions with different pseudo spins.
\par
\bigskip

\centerline{2. Equivalent, Antisymmetrized Variational Functionals}
\medskip
We will show that $ {\cal D} $ is the stationary value of either of the
following three Schwinger-like functionals of two independent trial
functions $ \Psi ,\Psi ^\prime , $
\smallskip
$$ {\cal F}=<\Psi ^\prime \mid{\cal A}V\mid \chi >+<\chi ^\prime \mid
V^\prime \mid \Psi >-<\Psi ^\prime \mid (z-H)\mid \Psi >, \eqno (2.1a) $$
$$ {\cal F}^\prime =<\Psi ^\prime \mid V\mid \chi >+<\chi ^\prime \mid
V^\prime{\cal A}\mid \Psi >-<\Psi ^\prime \mid (z-H)\mid \Psi >, \eqno (2.1b)
$$
$$ F=<\Psi ^\prime \mid{\cal A}V\mid \chi >+<\chi ^\prime \mid V^\prime{\cal
A}\mid \Psi >-<\Psi ^\prime \mid (z-H)\mid \Psi >, \eqno (2.1c) $$
\par
\noindent which differ by slightly different insertions of $ {\cal A}. $ The
variations of $ \Psi ,\Psi ^\prime $
give the following stationarity conditions,
\smallskip
$$ {\cal A}V\mid \chi >-(z-H)\mid \Psi >=0\ {\rm and} \ <\chi ^\prime \mid
V^\prime -<\Psi ^\prime \mid (z-H)=0\ {\rm for} \ {\cal F}, \eqno (2.2a) $$
$$ V\mid \chi >-(z-H)\mid \Psi >=0\ {\rm and} \ <\chi ^\prime \mid
V^\prime{\cal A}-<\Psi ^\prime \mid (z-H)=0\ {\rm for} \ {\cal F}^\prime ,
\eqno (2.2b) $$
$$ {\cal A}V\mid \chi >-(z-H)\mid \Psi >=0\ {\rm and} \ <\chi ^\prime \mid
V^\prime{\cal A}-<\Psi ^\prime \mid (z-H)=0\ {\rm for} \ F. \eqno (2.2c) $$
\par
\noindent For $ z $ complex, the full resolvent $ G\equiv (z-H)^{-1} $ is a
bounded, {\sl uniquely\/} defined
operator. Its action upon square integrable vectors $ V\mid \chi >, $ $ {\cal
A}V\mid \chi >, $ $ <\chi ^\prime \mid V^\prime , $
$ <\chi ^\prime \mid V^\prime{\cal A} $ returns also square integrable
vectors. Thus stationarity is
reached for a unique pair of trial functions,
\smallskip
$$ \mid \Psi >=G{\cal A}V\mid \chi >\ {\rm and} \ <\Psi ^\prime \mid =<\chi
^\prime \mid V^\prime G\ {\rm for} \ {\cal F}, \eqno (2.3a) $$
$$ \mid \Psi >=GV\mid \chi >\ {\rm and} \ <\Psi ^\prime \mid =<\chi ^\prime
\mid V^\prime{\cal A}G\ {\rm for} \ {\cal F}^\prime , \eqno (2.3b) $$
$$ \mid \Psi >=G{\cal A}V\mid \chi >\ {\rm and} \ <\Psi ^\prime \mid =<\chi
^\prime \mid V^\prime{\cal A}G\ {\rm for} \ F. \eqno (2.3c) $$
\par
\noindent The corresponding stationary value of each functional is then
\smallskip
$$ \widehat{{\cal F}}=(<\chi ^\prime \mid V^\prime G){\cal A}V\mid \chi
>+<\chi ^\prime \mid V^\prime (G{\cal A}V\mid \chi >)-(<\chi ^\prime \mid
V^\prime G)(z-H)(G{\cal A}V\mid \chi >), \eqno (2.4a) $$
$$ \widehat{{\cal F}^\prime} =(<\chi ^\prime \mid V^\prime{\cal A}G)V\mid
\chi >+<\chi ^\prime \mid V^\prime{\cal A}(GV\mid \chi >)-(<\chi ^\prime \mid
V^\prime{\cal A}G)(z-H)(GV\mid \chi >), \eqno (2.4b) $$
$$ \widehat{ F}=
(<\chi ^\prime \mid V^\prime{\cal A}G){\cal A}V\mid \chi
>+<\chi ^\prime \mid V^\prime{\cal A}(G{\cal A}V\mid \chi >)
$$
$$ \, \, \, \, \, \, \,
-(<\chi ^\prime
\mid V^\prime{\cal A}G)(z-H)(G{\cal A}V\mid \chi >), \eqno (2.4c) $$
\par
\noindent and the idempotence of $ {\cal A} $, together with its commutation
with $ H $ and $ G, $
reduce all nine terms in the right-hand sides (r.h.s.) of Eqs.(2.4) to the
same value, namely $ {\cal D}, $ see Eq.(1.1). Hence $ \widehat{{\cal F}}=
\widehat{{\cal F}^\prime} = \widehat{ F}={\cal D}. $ \par
\smallskip
We notice that six out of these nine terms are integrals confined to
an interaction volume, defined by the ranges of $ V\mid \chi >, $ $ <\chi
^\prime \mid V^\prime . $ For $ z $
reaching the on shell limit, $ \Psi ,\Psi ^\prime $ are no more square
integrable states, but
rather purely outgoing or ingoing waves. The calculation of $ {\cal D}(z), $
however,
by means of any one among these six terms, and the calculation of its on
shell limit $ {\cal D}(E+i0), $ can be confined to a truncation of $
\Psi ,\Psi ^\prime $ to this
interaction volume.  The remaining three terms contain both $\Psi$
and $\Psi^\prime$ and can not be used to calculate ${\cal D}(z)$
in the on-shell limit.    \par
\smallskip
For this reason, we now investigate an approximation in which $ \Psi ,\Psi
^\prime $
are replaced by products $ \Phi ,\Phi ^\prime $ of independent, square
integrable, single
particle orbitals $ \varphi_ i,\varphi^{ \prime}_ i, $ respectively, $ \Phi =
\prod^{ }_ i\varphi_ i, $ $ \Phi ^\prime = \prod^{ }_ i\varphi^{ \prime}_ i, $
{\sl or\/} antisymmetrized
products of such orbitals (Slater determinants). According to Eq.(2.3a), $
\Psi ^\prime $
contains only the same partial antisymmetrization as that contained in $ \chi
^\prime , $
and its approximation $ \Phi ^\prime $ should thus read as a product $ \Phi^{
\prime}_ c\Phi^{ \prime}_ d $ of separate
Slater determinants for the ejectile and the residual target degrees of
freedom, while $ \Psi $ is fully antisymmetrized, hence $ \Phi $ must be a
Slater
determinant. This approximation for $ {\cal F} $ thus reads
\smallskip
$$ {\cal F}=<\Phi^{ \prime}_ c\Phi^{ \prime}_ d\mid{\cal A}V\mid \chi >+<\chi
^\prime \mid V^\prime \mid \Phi >-<\Phi^{ \prime}_ c\Phi^{ \prime}_ d\mid
(z-H)\mid \Phi >. \eqno (2.5a) $$
\par
\noindent In the same way, the corresponding approximation for $ {\cal
F}^\prime $ reads
\smallskip
$$ {\cal F}^\prime =<\Phi ^\prime \mid V\mid \chi >+<\chi ^\prime \mid
V^\prime{\cal A}\mid \Phi_ a\Phi_ b>-<\Phi ^\prime \mid (z-H)\mid \Phi_
a\Phi_ b>, \eqno (2.5b) $$
\par
\noindent with partial antisymmetrization restricted to a and b. Finally for $
F $ both
$ \Psi ,\Psi ^\prime $ are fully antisymmetric, and the approximation for $ F
$ reads
\smallskip
$$ F=C+C^\prime -D,\ {\rm with} \ C=<\Phi ^\prime \mid V\mid \chi >,\
C^\prime =<\chi ^\prime \mid V^\prime \mid \Phi >,\ D=<\Phi ^\prime \mid
(z-H)\mid \Phi >, \eqno (2.5c) $$
\par
\noindent where the insertions of $ {\cal A} $ present in Eq.(2.1c) are
useless since $ \Phi ,\Phi ^\prime $ are
Slater determinants. \par
\smallskip
We now notice that the first term in the r.h.s.\ of Eq.(2.5a) and the
second term in the r.h.s.\ of Eq.(2.5b) contain an explicit $ {\cal A}, $ which
projects
products of Slater determinants $ \Phi_ a\Phi_ b, $ $ \Phi^{ \prime}_ c\Phi^{
\prime}_ d, $ into full determinants $ \Phi ,\Phi ^\prime , $
respectively. The same property belongs to the third term in the r.h.s.\ of
Eq.(2.5a), because the implicit $ {\cal A} $ contained in $ \Phi $ commutes
with $ H $ and
carries full antisymmetrization further, into the bra. A similar argument
holds again for the third term of the r.h.s.\ of Eq.(2.5b). It can be concluded
that $ {\cal F},{\cal F}^\prime ,F, $ which are equivalent when $ \Psi ,\Psi
^\prime $ are fully flexible trial
fuctions, are still equivalent when one attempts the separation of single
particle degrees of freedom. \par
\smallskip
The next Section, Sec.3, thus refers only to the variation of $ F, $ as
defined by Eq.(2.5c), with Slater determinants $ \Phi ,\Phi ^\prime $ as trial
functions. For
obvious technical reasons, we will investigate a slightly less general
problem than that announced when introducing Eq.(1.1): only those cases are
retained where $ \chi $ (resp. $ \chi ^\prime ) $ is the product of two
Slater determinants $ \chi_ a, $
$ \chi_ b, $ (resp. $ \chi^{ \prime}_ c, $ $ \chi^{ \prime}_ d). $ The wave
functions $ \chi_ a, $...$ \chi^\prime_d $ for the nuclei involved in
the collision correspond to boosted shell models, like those which define
the initial states of TDHF $ {\rm calculations} [3]. $ This technical
restriction still
defines a generic problem, however, since any matrix element of the
operator $ V^\prime{\cal A}GV $ between correlated internal states of these
nuclei can be
expanded in matrix elements between uncorrelated states, via usual, static
configuration mixings. \par
\bigskip

\centerline{3. Time Independent Mean Field Approximation}
\medskip
Let $ \left\{ \chi_ \alpha ,\alpha =1,...N_a \right\} $ be the set of
(boosted) orbitals which describes
the projectile. Let $ \left\{ \chi_ \beta ,\beta =N_a+1,...N \right\} , $ $
\left\{ \chi^\prime_\gamma ,\gamma =1,...N_c \right\} , $
 $ \left\{ \chi^\prime_\delta
,\delta =N_c+1,...N \right\} $ be
similar sets for nuclei b, c, d. It is convenient here to define the
following wave functions,
\smallskip
$$ \mid\bar  \chi >\ =\ \prod^{ N_a}_{\alpha =1}\mid \chi_ \alpha >\ \prod^
N_{\beta =N_a+1}\mid \chi_ \beta >,\ <\bar  \chi ^\prime \mid \ =\ \prod^{
N_c}_{\gamma =1}<\chi^{ \prime}_ \gamma \mid \ \prod^ N_{\delta
=N_c+1}<\chi^{ \prime}_ \delta \mid , \eqno (3.1a) $$
\par
\noindent which are pure products of orbitals, and
\smallskip
$$ \mid \widetilde{ \Phi} >= {\rm det} \left\{ \mid \varphi_ i> \right\} ,\ <
\widetilde{ \Phi} ^\prime \mid = {\rm det} \left\{ <\varphi^{ \prime }_
i\mid \right\} , \eqno (3.1b) $$
which are determinants of orbitals, without the usual $ (N!)^{-1/2} $
normalizations, because it is slightly easier to calculate the
contributions $ C,C^\prime $ to the functional $ F $ as
\smallskip
$$ C=(N_a!N_b!/N!)^{1/2}< \widetilde{ \Phi} ^\prime \mid V\mid\bar  \chi >,\
C^\prime =(N_c!N_d!/N!)^{1/2}<\bar  \chi ^\prime \mid V^\prime \mid
\widetilde{ \Phi} >. \eqno (3.1c) $$
\par
As usual, we consider the matrices of the overlaps of the single
particle orbitals, namely the matrices $ <\varphi^{ \prime}_ i\mid \chi_ j>, $
$ <\chi^{ \prime}_ i\mid \varphi_ j>, $ $ <\varphi^{ \prime}_ i\mid \varphi_
j> $ and their
inverse matrices $ A_{ij}, $ $ A^{\prime}_{ ij}, $ $ B_{ij}, $ respectively.
This generates the standard
cofactors $ M_{ij}, $ $ M_{ijlk}, $ $ M_{ijknml}, $...$ M^{\prime}_{ ij},
$...$ N_{ijknml} $ such as defined by
\smallskip
$$ < \widetilde{ \Phi} ^\prime \mid\bar  \chi >= {\rm det} \left(<\varphi^{
\prime}_ i\mid \chi_ j> \right)= \sum^{ }_ j<\varphi^{ \prime}_ i\mid \chi_
j>M_{ij} \ {\rm with} \ M_{ij}=< \widetilde{ \Phi} ^\prime \mid\bar  \chi
>A_{ji}, \eqno (3.2a) $$
$$ M^{\prime}_{ ik}= \sum^{ }_ l<\chi^{ \prime}_ j\mid \varphi_
l>M^{\prime}_{ ijlk} \ {\rm with} \ M^{\prime}_{ ijlk}=<\bar  \chi ^\prime
\mid \widetilde{ \Phi} >(A^{\prime}_{ ki}A^{\prime}_{ lj}-A^{\prime}_{
kj}A^{\prime}_{ li}), \eqno (3.2b) $$
$$ N_{ijml}= \sum^{ }_ n<\varphi^{ \prime}_ k\mid \varphi_n>N_{ijknml} \
{\rm with} \ N_{ijknml}=<\Phi ^\prime \mid \Phi > \times $$
$$
(B_{li}B_{mj}B_{nk}-B_{li}B_{mk}B_{nj}-B_{lj}B_{mi}B_{nk}+B_{lj}B_{mk}B_{ni}+\
B_{lk}B_{mi}B_{nj}-B_{lk}B_{mj}B_{ni}). \eqno (3.2c) $$
Accordingly, one obtains the derivatives
\smallskip
$$ {\partial <\Phi ^\prime \mid \Phi > \over \partial <\varphi^{ \prime}_
i\mid \varphi_ j>}=N_{ij}, \eqno (3.3a) $$
$$ {\partial M^{\prime}_{ kl} \over \partial <\chi^{ \prime}_ i\mid \varphi_
j>}=M^{\prime}_{ kijl}, \eqno (3.3b) $$
$$ {\partial M_{klnm} \over \partial <\varphi^{ \prime}_ i\mid \chi_
j>}=M_{klijnm}. \eqno (3.3c) $$
A straightforward calculation then gives
\smallskip
$$ < \widetilde{ \Phi} ^\prime \mid V\mid\bar  \chi >\ =\ {1 \over 2}\ \sum^{
}_{ ij\alpha \beta} <\varphi^{ \prime}_ i\varphi^{ \prime}_ j\mid v\mid \chi_
\alpha \chi_ \beta >M_{ij\beta \alpha} , \eqno (3.4a) $$
$$ <\bar  \chi ^\prime \mid V\mid \widetilde{ \Phi} >\ =\ {1 \over 2}\ \sum^{
}_{ \gamma \delta ij}<\chi^{ \prime}_ \gamma \chi^{ \prime}_ \delta \mid
v\mid \varphi_ i\varphi_ j>M^{\prime}_{ \gamma \delta ji}, \eqno (3.4b) $$
\par
\noindent$ <\Phi ^\prime \mid (z-H)\mid \Phi >=z {\rm det} (<\varphi^{
\prime}_ i\mid \varphi_ j>)\ - $
\smallskip
$$ \ \ \ \ \ \ \ \ \ \ \ \ \ \ \sum^{ }_{ ij}<\varphi^{ \prime}_ i\mid t\mid
\varphi_ j>N_{ij}-\ {1 \over 4}\ \sum^{ }_{ ijkl}<\varphi^{ \prime}_
i\varphi^{ \prime}_ j\mid v\mid \varphi_ k\varphi_ l>N_{ijlk}, \eqno (3.4c) $$
\par
\noindent where Latin indices $ i,j...l $ run from 1 to $ N, $ while Greek
indices $ \alpha ,\beta ,\gamma ,\delta $
run only inside the labels allowed for nuclei a,b,c,d, respectively. All
matrix elements of $ v $ are antisymmetrized. \par
\smallskip
The functional derivatives of $ C $ and $ C^\prime $ then read \par
\smallskip
$$ {\delta C \over \delta <\varphi^{ \prime}_ i\mid } = (N_a!N_b!/N!)^{1/2}\
\times $$
$$ \left( \sum^{ }_{ j\alpha \beta} <.\ \varphi^{ \prime}_ j\mid v\mid \chi_
\alpha \chi_ \beta >M_{ij\beta \alpha} +\ {1 \over 2}\ \sum^{ }_{ jmn\alpha
\beta} <\varphi^{ \prime}_ m\varphi^{ \prime}_ n\mid v\mid \chi_ \alpha \chi_
\beta >M_{mnij\beta \alpha} \mid \chi_ j> \right), \eqno (3.5a) $$
\par
$$ { \delta C^\prime \over \delta \mid \varphi_ j>}= (N_c!N_d!/N!)^{1/2}\
\times $$
$$ \left( \sum^{ }_{ i\gamma \delta} <\chi^{ \prime}_ \gamma \chi^{ \prime}_
\delta \mid v\mid . \ \varphi_ i>M^{\prime}_{ \gamma \delta ij}+\ {1 \over 2}\
\sum^{ }_{ i\gamma \delta mn}<\chi^{ \prime}_ \gamma \chi^{ \prime}_ \delta
\mid v\mid \varphi_ m\varphi_ n>M^{\prime}_{ \gamma \delta ijnm}<\chi^{
\prime}_ i\mid \right). \eqno (3.5b) $$
This suggests the definition of the following non Hermitian, channel mean
field potentials,
\smallskip
$$ <.\mid S_b\mid \chi_ \alpha >=< \widetilde{ \Phi} ^\prime \mid\bar  \chi
>^{-1} \sum^{ }_{ j\beta} <. \ \varphi^{ \prime}_ j\mid v\mid \chi_
\alpha \chi_
\beta >M_{j\beta} , \eqno (3.6a) $$
$$ <.\mid S_a\mid \chi_ \beta >=< \widetilde{ \Phi} ^\prime \mid\bar  \chi
>^{-1} \sum^{ }_{ j\alpha} <. \ \varphi^{ \prime}_ j\mid v\mid \chi_
\beta \chi_
\alpha >M_{j\alpha} , \eqno (3.6b) $$
$$ <\chi^{ \prime}_ \gamma \mid S_d^\prime \mid .>
=<\bar  \chi ^\prime \mid
\widetilde{ \Phi} >^{-1} \sum^{ }_{ \delta i}
<\chi^{ \prime}_ \gamma \chi^{
\prime}_ \delta \mid v\mid . \
\varphi_ i>M^{\prime}_{ \delta i}, \eqno (3.6c) $$
$$ <\chi^{ \prime}_ \delta \mid S_c^\prime \mid .>=<\bar  \chi ^\prime \mid
\widetilde{ \Phi} >^{-1} \sum^{ }_{ \gamma i}<\chi^{ \prime}_ \delta \chi^{
\prime}_ \gamma \mid v\mid . \
\varphi_ i>M^{\prime}_{ \gamma i}, \eqno (3.6d) $$
\par
\noindent which correspond to traces of $ v $ on "densities" related to nuclei
b,a,d,c,
respectively. It will be noticed that $ S_b $ acts upon orbitals $ \chi_
\alpha $ of nucleus a
only. In the same way, $ S_a, $ a trace on a "density" related to nucleus a,
acts upon orbitals $ \chi_ \beta $ of nucleus b only. Accordingly, both
formulae,
Eqs.(3.6a,b), can be interpreted as the definition of a unique mean field
potential $ S. $ Completely similar considerations are valid with
Eqs.(3.6c,d),
both traces on nuclei d and c defining a potential $ S^\prime , $ acting in
turn on
orbitals $ \chi^{ \prime}_ \gamma $ and $ \chi^{ \prime}_ \delta , $
respectively. \par
\smallskip
The functional derivatives of $ D $ are \par
\smallskip
$$ {\delta D \over \delta <\varphi^{ \prime}_ i\mid } = \sum^{ }_
j(z-t)\mid \varphi_ j>N_{ij}- \sum^{ }_{ jkl}<\varphi^{ \prime}_ k\mid t\mid
\varphi_ l>N_{kijl}\mid \varphi_ j>- $$
$$ {1 \over 2}\ \sum^{ }_{ jkl}<.\ \varphi^{ \prime}_ k\mid v\mid \varphi_
j\varphi_ l>N_{iklj}-\ {1 \over 4}\ \sum^{ }_{ jmnkl}<\varphi^{ \prime}_
m\varphi^{ \prime}_ n\mid v\mid \varphi_ k\varphi_ l>N_{mnijlk}\mid \varphi_
j>, \eqno (3.5c) $$
\par
$$ {\delta D \over \delta \mid \varphi_ j>}= \sum^{ }_i<\varphi_i^\prime \mid
(z-t)N_{ij}- \sum^{ }_{ ikl}<\varphi^{ \prime}_ k\mid t\mid \varphi_
l>N_{kijl}<\varphi^{ \prime}_ i\mid - $$
$$ {1 \over 2}\ \sum^{ }_{ ikl}<\varphi^{ \prime}_ i\varphi^{ \prime}_ k\mid
v\mid .\ \varphi_ l>N_{iklj}-\ {1 \over 4}\ \sum^{ }_{ imnkl}<\varphi^{
\prime}_ m\varphi^{ \prime}_ n\mid v\mid \varphi_ k\varphi_
l>N_{mnijlk}<\varphi^{ \prime}_ i\mid . \eqno (3.5d) $$
\par
This suggests the definition of the following non Hermitian, mean
field potential,
\smallskip
$$ <.\mid U\mid \varphi_ j>=<\Phi ^\prime \mid \Phi >^{-1} \sum^{ }_{
kl}<. \ \varphi^{ \prime}_ k\mid v\mid \varphi_ j\varphi_ l>N_{kl}, \eqno
(3.6e)
$$
$$ <\varphi^{ \prime}_ i\mid U\mid .>=<\Phi ^\prime \mid \Phi >^{-1} \sum^{
}_{ kl}<\varphi^{ \prime}_ i\varphi^{ \prime}_ k\mid v\mid . \
\varphi_ l>N_{kl}.
\eqno (3.6f) $$
With such definitions of $ S,S^\prime ,U, $ the functional derivatives now
read \par
$$ {\delta C \over \delta <\varphi^{ \prime}_ i\mid } = (N_a!N_b!/N!)^{1/2}<
\widetilde{ \Phi} ^\prime \mid\bar  \chi > \times $$
$$ \left[ \sum^{ }_ j \left(S+ \sum^{ }_{ kl}<\varphi^{ \prime}_ k\mid{ S
\over 2}\mid \chi_ l>A_{lk} \right)\mid \chi_ j>A_{ji}- \sum^{ }_{
jkl}<\varphi^{ \prime}_ k\mid S\mid \chi_ l>A_{li}A_{jk}\mid \chi_ j> \right],
\eqno (3.7a) $$
\par
$$ {\delta C^\prime \over \delta \mid \varphi_ j>} =
(N_c!N_d!/N!)^{1/2}<\bar  \chi ^\prime \mid \widetilde{ \Phi} > \times $$
$$ \left[ \sum^{ }_ iA^{\prime}_{ ji}<\chi^{ \prime}_ i\mid \left(S^\prime +
\sum^{ }_{ kl}<\chi^{ \prime}_ k\mid{ S^\prime \over 2}\mid \varphi_
l>A^{\prime}_{ lk} \right)- \sum^{ }_{ ikl}<\chi^{ \prime}_ k\mid S^\prime
\mid \varphi_ l>A^{\prime}_{ li}A^{\prime}_{ jk}<\chi^{ \prime}_ i\mid
\right]. \eqno (3.7b) $$
\par
$$ {\delta D \over \delta <\varphi^{ \prime}_ i\mid}  = <\Phi
^\prime \mid \Phi >\ \left[ \sum^{ }_ j \left(z-t-U- \sum^{ }_{ kl}<\varphi^{
\prime}_ k\mid (t+{U \over 2})\mid \varphi_ l>B_{lk} \right)\mid \varphi_
j>B_{ji}+ \right. $$
$$ \left. \sum^{ }_{ jkl}<\varphi^{ \prime}_ k\mid (t+U)\mid \varphi_
l>B_{li}B_{jk}\mid \varphi_ j> \right], \eqno (3.7c) $$
$$ {\delta D \over \delta \mid \varphi_ j>} = <\Phi ^\prime \mid
\Phi >\ \left[ \sum^{ }_ iB_{ji}<\varphi^{ \prime}_ i\mid \left(z-t-U- \sum^{
}_{ kl}<\varphi^{ \prime}_ k\mid (t+{U \over 2})\mid \varphi_ l>B_{lk}
\right)+ \right. $$
$$ \left. \sum^{ }_{ ikl}<\varphi^{ \prime}_ k\mid (t+U)\mid \varphi_
l>B_{li}B_{jk}<\varphi^{ \prime}_ i\mid \right]. \eqno (3.7d) $$
\par
We now take advantage of the fact that a Slater determinant is
invariant, except for an inessential change of phase and norm, under the
linear rearrangement of its orbitals. Hence a linear rearrangement of
orbitals $ \varphi_ j $ can diagonalize one out the four matrices $
<\varphi^{ \prime}_ i\mid \varphi_ j>, $ $ <\varphi^{ \prime}_ i\mid h\mid
\varphi_ j>, $
$ <\chi^{ \prime}_ i\mid \varphi_ j>, $ $ <\chi^{ \prime}_ i\mid S^\prime
\mid \varphi_ j>. $ In the same way, a linear rearrangement of orbitals $
\varphi^{ \prime}_ i $
can diagonalize one out of the four matrices $ <\varphi^{ \prime}_ i\mid
\varphi_ j>, $ $ <\varphi^{ \prime}_ i\mid h\mid \varphi_ j>, $ $ <\varphi^{
\prime}_ i\mid \chi_ j>, $
$ <\varphi^{ \prime}_ i\mid S\mid \chi_ j>. $
In a representation where both overlap matrices $ <\varphi^{ \prime}_ i\mid
\chi_ j> $ and $ <\chi^{ \prime}_ i\mid \varphi_ j> $
are diagonal, these formulae reduce to
$$ {\delta C \over \delta <\varphi^{ \prime}_ i\mid} = (N_a!N_b!/N!)^{1/2}<
\widetilde{ \Phi} ^\prime \mid\bar  \chi > <\varphi^{ \prime}_ i\mid \chi_
i>^{-1}\ \times $$
$$ \left[ \left(S+{1 \over 2} \sum^{ }_ k{<\varphi^{ \prime}_ k\mid S\mid
\chi_ k> \over <\varphi^{ \prime}_ k\mid \chi_ k>} \right)\mid \chi_ i>-
\sum^{ }_ j{<\varphi^{ \prime}_ j\mid S\mid \chi_ i> \over <\varphi^{
\prime}_ j\mid \chi_ j>}\mid \chi_ j> \right], \eqno (3.8a) $$
$$ {\delta C^\prime \over\delta \mid \varphi_ j> } =
(N_c!N_d!/N!)^{1/2}<\bar  \chi ^\prime \mid \widetilde{ \Phi} > <\chi^{
\prime}_ j\mid \varphi_ j>^{-1}\ \times $$
$$ \left[<\chi^{ \prime}_ j\mid \left(S^\prime +{1 \over 2} \sum^{ }_
k{<\chi^{ \prime}_ k\mid S^\prime \mid \varphi_ k> \over <\chi^{ \prime}_
k\mid \varphi_ k>} \right)- \sum^{ }_ i{<\chi^{ \prime}_ j\mid S^\prime \mid
\varphi_ i> \over <\chi^{ \prime}_ i\mid \varphi_ i>}<\chi^{ \prime}_ i\mid
\right], \eqno (3.8b) $$
where one finds that the reference energy for $ S, $ resp. $
S^\prime , $ is
$ <\Phi ^\prime \mid V\mid \chi >/<\Phi ^\prime \mid \chi >, $ resp. $ <\chi
^\prime \mid V^\prime \mid \Phi >/<\chi ^\prime \mid \Phi >, $
see also Eq.(3.9).  Alternately, in
a representation
where both matrices $ <\varphi^{ \prime}_ i\mid \varphi_ j> $ and $
<\varphi^{ \prime}_ i\mid h\mid \varphi_ j> $, with $ h\equiv t+U, $ are
diagonal, the
simplification reads
\smallskip
$$ {\delta D \over \delta <\varphi^{ \prime}_ i\mid} ={<\Phi ^\prime \mid
\Phi > \over <\varphi^{ \prime}_ i\mid \varphi_ i>} \left(z-h- \sum^{ }_
k{<\varphi^{ \prime}_ k\mid (t+{U \over 2})\mid \varphi_ k> \over <\varphi^{
\prime}_ k\mid \varphi_ k>}+{<\varphi^{ \prime}_ i\mid h\mid \varphi_ i>
\over <\varphi^{ \prime}_ i\mid \varphi_ i>} \right)\mid \varphi_ i>, \eqno
(3.8c) $$
$$ {\delta D \over \delta \mid \varphi_ j>}={<\Phi ^\prime \mid \Phi > \over
<\varphi^{ \prime}_ j\mid \varphi_ j>}<\varphi^{ \prime}_ j\mid \left(z-h-
\sum^{ }_ k{<\varphi^{ \prime}_ k\mid (t+{U \over 2})\mid \varphi_ k> \over
<\varphi^{ \prime}_ k\mid \varphi_ k>}+{<\varphi^{ \prime}_ j\mid h\mid
\varphi_ j> \over <\varphi^{ \prime}_ j\mid \varphi_ j>} \right), \eqno (3.8d)
$$
\par
\noindent where one finds the reference energy for $ h, $
\smallskip
$$ \bar  E\ =\ {<\Phi ^\prime \mid H\mid \Phi > \over <\Phi ^\prime \mid \Phi
>}\ =\ \sum^{ }_ k{<\varphi^{ \prime}_ k\mid (t+{U \over 2})\mid \varphi_ k>
\over <\varphi^{ \prime}_ k\mid \varphi_ k>}, \eqno (3.9) $$
\par
\noindent and self energies for $ \varphi_ i,\varphi^{ \prime}_ i, $
\smallskip
$$ \eta_ i\ =\ z-\ \bar  E\ +\ {<\varphi^{ \prime}_ i\mid h\mid \varphi_ i>
\over <\varphi^{ \prime}_ i\mid \varphi_ i>}. \eqno (3.10) $$
\par
 We notice that the unknown
orbitals $ \varphi_ i $ are much less coupled to
one another in Eq.(3.8c) than in Eq.(3.7c), and the same remark holds for
the $ \varphi^{ \prime}_ j $ when Eq.(3.7d) simplifies into Eq.(3.8d). Hence,
among all the
possible rearrangements which could technically simplify practical
calculations, we choose those representations which simultaneously
diagonalize $ <\varphi^{ \prime}_ i\mid \varphi_ j> $ and $ <\varphi^{
\prime}_ i\mid h\mid \varphi_ j>. $   The remaining four matrices
will in general be non diagonal.
Eqs.(3.8c), (3.8d) also have the advantage that they are close to the
traditional, homogeneous Hartree-Fock scheme.
 \par
\smallskip
\ \smallskip
The stationarity conditions,
$$ {\delta (C-D) \over \delta< \varphi^{ \prime }_
i \mid} =0 \, , \,  {\delta (C^\prime -D) \over \delta \mid \varphi_ j>}=0
\eqno (3.11) $$
 then read

\smallskip
$$ (\eta_ i-h)\mid \varphi_ i>\ =\ {(N_a!N_b!/N!)^{1/2}\ <
\widetilde{ \Phi} ^\prime \mid\bar  \chi >\ <\varphi^{ \prime}_ i\mid
\varphi_ i> \over <\Phi ^\prime \mid \Phi >}\ \times $$
\smallskip
$$ \left[ \sum^{ }_ j \left(S+ \sum^{ }_{ kl}<\varphi^{ \prime}_ k\mid{ S
\over 2}\mid \chi_ l>A_{lk} \right)\mid \chi_ j>A_{ji}- \sum^{ }_{
jkl}<\varphi^{ \prime}_ k\mid S\mid \chi_ l>A_{li}A_{jk}\mid \chi_ j> \right],
\eqno (3.12a) $$
\par
\noindent and \par
\smallskip
$$ <\varphi^{ \prime}_ j\mid (\eta_ j-h)=\ {(N_c!N_d!/N!)^{1/2}\
<\bar  \chi ^\prime \mid \widetilde{ \Phi} >\ <\varphi^{ \prime}_ j\mid
\varphi_ j> \over <\Phi ^\prime \mid \Phi >}\ \times $$
\smallskip
$$ \left[ \sum^{ }_ iA^{\prime}_{ ji}<\chi^{ \prime}_ i\mid \left(S^\prime +
\sum^{ }_{ kl}<\chi^{ \prime}_ k\mid{ S^\prime \over 2}\mid \varphi_
l>A^{\prime}_{ lk} \right)- \sum^{ }_{ ikl}<\chi^{ \prime}_ k\mid S^\prime
\mid \varphi_ l>A^{\prime}_{ li}A^{\prime}_{ jk}<\chi^{ \prime}_ i\mid
\right]. \eqno (3.12b) $$
\par
\noindent A slight simplification of the r.h.s.\ of Eqs.(3.12) is further
possible:
because $ \chi_ a $ and $ \chi_ b $ are Slater determinants, a linear
rearrangement of the
orbitals $ \chi_ \alpha $ among themselves, and a separate rearrangement of
the orbitals
$ \chi_ \beta , $ can partly diagonalize the matrix $ <\varphi^{ \prime}_
i\mid \chi_ j>, $ but the simplicity of
Eq.(3.8a) cannot be obtained under such a partial diagonalization. The same
holds for separate rearrangements of $ \chi^{ \prime}_ \gamma $ and $ \chi^{
\prime}_ \delta , $ which can induce the
simplified r.h.s., Eq.(3.8b), partly only.
\par
\smallskip
It will be interesting, in future numerical applications, to
investigate whether the residual off-diagonal parts of $ <\varphi^{ \prime}_
i\mid \chi_ j> $ and $ <\chi^{ \prime}_ i\mid \varphi_ j> $
can be neglected. If so, the corresponding approximation would create a
representation where all four equations, Eqs.(3.8), would be almost
compatible. This "one-orbital-partner" approximation would represent the
many-body collision by single-particle transitions $ \chi_ i\rightarrow
\varphi^{ \prime}_ i, $ $ \varphi_ i\rightarrow \chi^{ \prime}_ i, $ with a
conservation of the label $ i. $ \par
\smallskip
For the sake of completeness, we recall here a result, shown
$ {\rm earlier} [4] $ in a slightly different context where both $ \chi , $ $
\chi ^\prime $ are full Slater
determinants rather than products of such determinants. The result is still
valid for the present paper: \par
\noindent{\sl Theorem\/}: The solutions $ \varphi_ i, $ $ \varphi^{ \prime}_ j
$ of Eqs.(3.12) are consistent with the ansatz
that both matrices $ <\varphi^{ \prime}_ i\mid \varphi_ j> $ and $ <\varphi^{
\prime}_ i\mid h\mid \varphi_ j> $ are diagonal. \par
\noindent{\sl Proof\/}: For $ i\not= j, $ multiply Eq.(3.12a) by $ <\varphi^{
\prime}_ j\mid $ and Eq.(3.12b) by $ \mid \varphi_ i>, $
respectively. Both resulting r.h.s.\ vanish, because the antisymmetries of $
\Phi
^\prime $
and $ \Phi $ with respect to their respective orbitals induce
 $$ <\varphi^{
\prime}_ j\mid{ \delta C \over \delta <\varphi^{ \prime}_ i\mid} >=0 \,
 , \,
 <{\delta C^\prime \over \delta \mid \varphi_ j>}\mid \varphi_ i>=0.
\eqno (3.13) $$
Variation of $C$ and $C^\prime$ with respect to $<\varphi_i^\prime\mid$
and $\mid \varphi_j>$ and subsequent scalar multiplication by
$<\varphi_j^\prime\mid$ and $\mid \varphi_i>$ will for $j \neq i$
create
two identical rows ( or columns ) in the determinantal structure of
$C$ and $C^\prime$.
\bigskip

\centerline{4. Self Consistent Symmetries}
\medskip
In practical calculations one often uses channel wave functions
which have some symmetry. Then one can choose trial functions
having the same symmetry, without destroying the self consistency
of the mean field equations, if some commutation relations are
 fulfilled.
Using such self consistent symmetries
could be called a "projection before variation" and may reduce the number
of coupled mean field equations enormously.

Let us first consider a unitary operator $P$ acting in single-particle space,
which relates the
$\chi_i$'s through
$$ P|\chi_i > = e^{i \theta_i} |\chi_{\sigma (i)} > \, {\rm and } \,
  <\chi_i^\prime | P^{\dag}  = <\chi_{\sigma (i)}^\prime | e^{-i \theta_i} ,
\eqno (4.1)
$$
where $\sigma $ indicates a permutation among the orbitals
and $\theta_i $ is an arbitrary $i $-dependent phase.
Then one can choose
$$ P|\varphi_i > = e^{i \theta_i} |\varphi_{\sigma (i)} > \, {\rm and} \,
  <\varphi_i^\prime | P^{\dag}  = <\varphi_{\sigma (i)}^\prime | e^{-i
\theta_i}
 ,
\eqno (4.2)
$$
if the following commutation relations are valid
$$
[t,P] = 0 \, , \, [v,P \otimes P ]=0 ,
\eqno (4.3a)
$$
and if furthermore the transformation $P$ leaves the orbitals
in the same fragment or transfers all orbitals from one fragment to
the other, as we shall prove below.

For the overlap matrices
$\beta_{ij} = <\varphi_i^\prime | \varphi_j >, $
$\alpha_{ij} = < \varphi_i^\prime | \chi_j >, $
$\alpha_{ij}^\prime  = <\chi_i^\prime | \varphi_j > $,
the unitarity of $P$ implies
$$
\beta_{ij} = e^{i(\theta_j-\theta_i)} \beta_{\sigma (i) \sigma (j)}  \, , \,
\alpha_{ij} =  e^{i(\theta_j-\theta_i)} \alpha_{\sigma (i) \sigma (j)} \, , \,
\alpha_{ij}^\prime = e^{i(\theta_j-\theta_i)}
\alpha_{\sigma (i) \sigma (j)}^\prime .
\eqno (4.4)
$$
Next let us study the transformation properties of cofactors and
inverse matrices.  Writing the generalized cofactor expansion
of $\det \beta$, which is invariant under Eq.(4.2), as
$$ \delta_{ki} \det \beta =
   \sum_j \beta_{ij} N_{kj} =
   \sum_j e^{i(\theta_j-\theta_i)} \beta_{\sigma (i) \sigma (j)} N_{kj} ,
\eqno (4.5) $$
and comparing with
$$ \delta_{\sigma (k) \sigma (i)} \det \beta =
   \sum_j \beta_{\sigma (i)j} N_{\sigma (k) j} =
   \sum_j \beta_{\sigma (i) \sigma (j)} N_{\sigma (k) \sigma (j)} ,
\eqno (4.6) $$
one finds :
$$ N_{ij} = e^{-i(\theta_j-\theta_i)} N_{\sigma (i) \sigma (j) },
\eqno (4.7a) $$
and for the inverse of the overlap matrix $\beta $
$$ B_{ij} = e^{i(\theta_j-\theta_i)} B_{\sigma (i) \sigma (j) }.
\eqno (4.7b) $$
In the same way one gets
$$ M_{ij} = e^{-i(\theta_j-\theta_i)} M_{\sigma (i) \sigma (j) } \, , \,
   M^\prime_{ij}
    = e^{-i(\theta_j-\theta_i)} M^\prime_{\sigma (i) \sigma (j) } .
\eqno (4.7c) $$
Eq.(4.2) together with Eq.(4.3a) give
$$
<\varphi_i^\prime |t|\varphi_j> =
 e^{i(\theta_j-\theta_i)}
  <\varphi_{\sigma (i)}^\prime |t|\varphi_{\sigma (j)}> .
\eqno (4.8a) $$
and
$$
<\varphi_i^\prime \varphi_k^\prime |v|\varphi_j \varphi_l >=
 e^{i(\theta_j+\theta_l-\theta_i-\theta_k)}
 <\varphi_{\sigma (i)}^\prime \varphi_{\sigma (k)}^\prime |
v|\varphi_{\sigma (j)} \varphi_{\sigma (l)} >.
\eqno (4.8b) $$
{}From (4.8a,b) together with (4.7a,b) we find
for the self consistent one-body operators $U$ and $h$
$$
<\varphi_i^\prime |U|\varphi_j>=
 e^{i(\theta_j-\theta_i)}
<\varphi_{\sigma (i)}^\prime |U|\varphi_{\sigma (j)}>
\, {\rm and} \,
<\varphi_i^\prime |h|\varphi_j>=
 e^{i(\theta_j-\theta_i)}
<\varphi_{\sigma (i)}^\prime |h|\varphi_{\sigma (j)}> ,
\eqno (4.9a) $$
or extending to the full Hilbert space of single-particle motion
$$ [U,P] =0 \,  ,  \, [h,P] =0 .
\eqno (4.9b) $$
Eqs.(4.9a) and (4.9b) together with Eq.(4.7b) will prove sufficient
for the l.h.s.\ of the mean field equations (3.12) to be invariant.
We will now turn to the inhomogeneous terms on their r.h.s..

To this end, let us define  $ \rho $ (resp. $\rho^\prime  $) as a
matrix with elements equal to $0$ if the orbitals $\chi_i$ and $\chi_j$
(resp. $\chi_i^\prime $ and $\chi_j^\prime $)
are from the same fragment and $1$ if they are from different fragments.
Then if $P$ satisfies
$$
\rho_{ij}  = \rho_{\sigma (i) \sigma (j) }
\, \, {\rm and}  \,
\rho^\prime_{ij}= \rho^\prime_{\sigma (i) \sigma (j) } ,
\eqno (4.3b)  $$
we have
$$ < \cdot \ |[S,P]|\chi_i> = 0 \, , \,
<\chi_i^\prime |[S^\prime ,P] |\  \cdot >= 0 ,
\eqno (4.10a) $$
where
$$ S |\chi_i > = (\det\alpha )^{-1}\sum_{kl} \rho_{il}
< \cdot \ \varphi_k^\prime | v | \chi_i \chi_l > M_{kl} ,
\eqno (4.10b) $$
$$ < \chi_i^\prime | S^\prime = (\det\alpha^\prime )^{-1}\sum_{kl}
 \rho_{ki}^\prime
< \chi_i^\prime \chi_k^\prime | v | \cdot \varphi_l >
M_{kl}^\prime ,
\eqno (4.10c) $$
use having been made of Eq.(4.7c) and the analogue of Eq.(4.8b) for
the channel potentials.

To prove the self consistency
of the ansatz (4.2), one has to show that
$$ P {\delta F \over {\delta <\varphi_i^\prime|}} =
     e^{i\theta_i}
     {\delta F \over \delta {<\varphi^\prime_{\sigma (i)} | } } \ .
\eqno (4.11) $$
This is exemplified for Eq. (3.7c)
where the symbol $\bar \eta $ is used for the $i$-independent part of the
self energy
$$ P \det \beta \left[ \sum_j (z-h-\bar\eta) B_{ji}
+\sum_{jkl} <\varphi_k^\prime |h|\varphi_l > B_{li} B_{jk} \right]
|\varphi_j> =
 $$
$$ e^{i\theta_i } \det \beta \left[ \sum_j (z-h-\bar\eta)
 B_{\sigma (j) \sigma (i) }
+\sum_{jkl} <\varphi_{\sigma (k)}^\prime |h|\varphi_{\sigma (l)} >
B_{\sigma (l) \sigma (i)} B_{\sigma (j) \sigma (k)} \right]
|\varphi_{\sigma (j)}> ,
\eqno (4.12) $$
using (4.7b) and (4.9a,b). This is the desired result
because in the sum $ \sigma (j) $ can be
replaced by $j$ for every summation index. The same can be done for the
r.h.s.\  of the mean field equations (3.7a) and for Eqs.(3.7b,d),
which completes the proof of the self consistency of the ansatz (4.2).

The second case are antiunitary symmetries like time reversal.
Let us recall the properties of antiunitary operators $A$,
$$ A(\lambda |\varphi >)= \lambda^* A |\varphi> ,
\eqno (4.13a) $$
$$
<\varphi^\prime | ( A|\varphi > )= (<\varphi^\prime |A )|\varphi >^*
\eqno (4.13b) $$
\noindent
and
$$
A^{\dag}A=AA^{\dag}=1 \,\, {\rm or} \,
<\varphi^\prime |\varphi > =
\left[ (<\varphi^\prime | A^{\dag} )(A\varphi > ) \right]^* .
\eqno (4.13c) $$
Let us consider a situation where
$$ A |\chi_i> = e^{i\theta_i}|\chi^\prime_{\sigma (i)} >
\, {\rm and} \,
   <\chi_i^\prime | A^{\dag} = <\chi_{\sigma (i)} | e^{-i\theta_i} .
\eqno (4.14) $$
Then one can choose
$$
A |\varphi_i > = e^{i\theta_i}|\varphi^\prime_{\sigma (i) } >
\, {\rm and} \,
<\varphi_i^\prime | A^{\dag} = < \varphi_{\sigma (i)} | e^{-i\theta_i} ,
\eqno (4.15) $$
to fulfill the mean field equations provided $A$ obeys the following
 relations
$$
 A \, t=t^{\dag}A \, , \, (A \otimes A)v = v^{\dag} (A \otimes A)  ,
\eqno (4.16a)
$$
and leaves the orbitals in the same fragment or transfers all
orbitals from one fragment to the other such that
$$
\rho_{ij} = \rho_{\sigma (j) \sigma (i)}^\prime  \, , \,
\rho_{ij}^\prime = \rho_{\sigma (j) \sigma (i)} .
\eqno (4.16b) $$
For Hermitian $t$ and $v$, Eqs.(4.16a)  reduce to commutator relations.
{}From (4.13c) it can be seen that
$$
\beta_{ij} = e^{i(\theta_i -\theta_j)} \beta_{\sigma (j) \sigma (i)}
\, ,\,
\alpha_{ij} = e^{i(\theta_i -\theta_j)} \alpha_{\sigma (j) \sigma (i)}^\prime
\, ,\,
\alpha_{ij}^\prime = e^{i(\theta_i -\theta_j)} \alpha_{\sigma (j) \sigma (i)}
\, \eqno (4.17a)
$$
and
$$
  \det \alpha = \det \alpha^\prime .
\eqno (4.17b)
$$
Like in (4.5) and (4.6), one has for example
$$
\delta_{ki} \det \alpha =  \sum_j \alpha_{ij} M_{kj}
=\delta_{\sigma (k) \sigma (i)} \det \alpha^\prime =
\sum_j e^{i(\theta_i -\theta_j)} \alpha^\prime_{\sigma (j) \sigma (i)}
 M^\prime_{\sigma (j) \sigma (k)} ,
\eqno (4.18) $$
hence we can conclude
$$
M_{ij} = e^{-i(\theta_i -\theta_j)} M_{\sigma (j) \sigma (i)}^\prime
\, , \,
M_{ij}^\prime = e^{-i(\theta_i -\theta_j)} M_{\sigma (j) \sigma (i)},
\eqno (4.19a)
$$
similarly
$$
N_{ij} = e^{-i(\theta_i -\theta_j)} N_{\sigma (j) \sigma (i)}
\, , \,
B_{ij} = e^{i(\theta_i -\theta_j)} B_{\sigma (j) \sigma (i)}.
\eqno (4.19b)
$$ For the matrix elements of $t$ one calculates $$ <\varphi^\prime_i
|t| \varphi_j> = e^{i(\theta_i - \theta_j)} <\varphi_{\sigma
(j)}^\prime |t|\varphi_{\sigma (i)} > ,
\eqno (4.20)
$$ while for $v$ $$ <\varphi^\prime_i \varphi^\prime_k | v| \varphi_j
\varphi_l > = < \varphi^\prime_i \varphi^\prime_k | \left[ v (A
\otimes A)^{\dag} (A \otimes A) | \varphi_j \varphi_l > \right] = $$
$$ \left( \left[ < \varphi^\prime_i \varphi^\prime_k | (A \otimes
A)^{\dag} \right] v^{\dag} \left[ (A \otimes A) | \varphi_j \varphi_l
> \right] \right)^* = e^{i(\theta_i + \theta_k - \theta_j -\theta_l)}
<\varphi_{\sigma (j)}^\prime \varphi_{\sigma (l)}^\prime |
v|\varphi_{\sigma (i)} \varphi_{\sigma (k)} > .
\eqno (4.21)
$$ Therefore $$ <\varphi^\prime_i |U| \varphi_j> = e^{i(\theta_i -
\theta_j)} <\varphi_{\sigma (j)}^\prime |U|\varphi_{\sigma (i)} >
\, , \,
<\varphi^\prime_i |h| \varphi_j> = e^{i(\theta_i - \theta_j)}
<\varphi_{\sigma (j)}^\prime |h|\varphi_{\sigma (i)} >,
\eqno (4.22)
$$ whereas $S$ is related to $S^\prime$ $$ <\varphi^\prime_i |S|
\chi_j> = e^{i(\theta_i - \theta_j)} <\chi_{\sigma (j)}^\prime
|S^\prime|\varphi_{\sigma (i)} >
\, , \,
<\chi^\prime_i |S^\prime | \varphi_j> = e^{i(\theta_i - \theta_j)}
<\varphi_{\sigma (j)}^\prime |S |\chi_{\sigma (i)} > .
\eqno (4.23) $$
Now we are able to express the commutation relations for $S$,
$S^\prime$ and $U$, $h$ with $A$ as $$ AU = U^{\dag}A \, , \,
Ah=h^{\dag}A ,
\eqno (4.24a) $$
$$ < \cdot \ |( AS | \chi_i >)= < \cdot \ |(S^{\prime \dag}A | \chi_i >)
\,  , \,
(< \chi_i^\prime |AS^\prime )|\ \cdot > = (<\chi_i^\prime |S^{\dag} A
)|\ \cdot > ,
\eqno (4.24b) $$
and as in (4.12) we can conclude $$ A \det \beta \left[ \sum_j
(z-h-\bar \eta )B_{ji} + \sum_{jkl} <\varphi_k^\prime |h| \varphi_l >
B_{li} B_{jk} \right] |\varphi_j> = $$ $$ e^{i\theta_i} \det \beta
\left[ \sum_j (z^*-h^{\dag} -{\bar \eta}^* )B_{\sigma(i)j}^* + \sum_{jkl}
<\varphi_k^\prime |h| \varphi_l >^* B_{lj}^* B_{\sigma (i) k}^*
\right] |\varphi_j^\prime> \, .
\eqno (4.25)
$$ Apart from a phase factor we recognize the l.h.s.\ of the complex
conjugate mean field equation resulting from $\delta F / \delta
|\varphi_{\sigma(i)}>$.  Doing the same for the r.h.s.\ proves the
consistency of the ansatz (4.15).  It is worth noting that the above
proofs simplify if one uses the diagonal representation in
channel-spin formalism (see Appendix).

It is possible to generalize (4.1) or (4.14).  We consider the case
where the unitary operator $P$ acts only inside the given set of
orbitals $\chi_i$ and $\chi_i^\prime$ such that
$$
P |\chi_i > =
\sum_j L_{ji} |\chi_j>
\, {\rm and} \,
<\chi_i^\prime | P^{\dag} = \sum_j <\chi^\prime_j |L^{\dag}_{ij} \ \ .
\eqno (4.26)
$$
Then the following ansatz is self consistent
$$
P |\varphi_i > =
\sum_j L_{ji} |\varphi_j>
\, {\rm and} \,
<\varphi_i^\prime | P^{\dag} = \sum_j <\varphi^\prime_j | L^{\dag}_{ij}\ \ ,
\eqno (4.27)
$$
i.e. it fulfills
$$
P {\delta F \over {\delta <\varphi_i^\prime|}} =
\sum_j L_{ji} {\delta F \over \delta { <\varphi^\prime_j | } } =
{\delta F \over \delta {( <\varphi^\prime_i |P^+ ) } } \ \ ,
\eqno (4.28)
$$
provided the commutator conditions (4.3a) are satisfied together with
$$
[\rho , L] = [\rho^\prime , L ] = 0 .
\eqno (4.29)
$$
If (4.28) is valid then $P$ generates out of the $i$-th mean field
equation the equation for the transformed orbital, which therefore can
be left out of the actual calculation.  The proof starts by noting
that
$$
\beta_{ij} = \sum_{kl} L^{\dag}_{ik} L_{lj} \beta_{kl} ,
\eqno (4.30a)
$$
in short
$$
\beta = L^{\dag} \beta L ,
\eqno (4.30b)
$$
which leads to
$$
[\beta ,L] = 0 \, , \ [\tilde N,L] =0 \ \ {\rm and} \ \ [B, L] = 0 .
\eqno (4.31a)
$$
Here $\tilde N$ is the transpose of the cofactor matrix $N_{ij}$.
By the same argument
$$
[\tilde M,L]=0 \, {\rm and} \, \ [\tilde M^\prime,L]=0 .
\eqno (4.31b)
$$
Using (4.3a) one finds
$$
U \det \beta =\sum_{kl} < \cdot \  \varphi^\prime_k|v|\cdot \varphi_l> N_{kl}
= P^{\dag} \left( \sum_{kl} \sum_{mn} L^{\dag}_{kn} L_{ml}
< \cdot \ \varphi^\prime_k|v|\cdot \ \varphi_l> N_{kl} \right) P ,
\eqno (4.32)
$$
and applying (4.31a) results in
$$
 U = P^{\dag} \left( (\det \beta )^{-1}
\sum_{mn} N_{nm} < \cdot \ \varphi^\prime_n|v|\cdot \varphi_m> \right) P
=P^{\dag}UP.
\eqno (4.33)
$$
Thus one can list the following properties
$$
[U,P]=0 \, , \, [S,P]=[S^\prime ,P]=0 \, {\rm and } \, [h,P]=0 ,
\eqno (4.34)
$$
which lead to
$$
[\bar U,L]=0 \, , \, [\bar S,L]=[\bar S^\prime ,L]=0\, , \,
\, {\rm and } \, [\bar h,L]=0 .
\eqno (4.35)
$$
$\bar U$, $\bar S$, $\bar S^\prime $  and $\bar h$ are defined as the matrices
$U_{ij}=<\varphi_i^\prime |U|\varphi_j>$ etc.
With these relations the proof of (4.28) is simple.
For the example (3.7c) the l.h.s.\ of (4.28) can be written as
$$
(\det \beta ) \sum_j \left[ (z-h-\bar \eta ) \left( L B \right)_{ji}
 + \left( LB\bar hB \right)_{ji} \right] |\varphi_j >
=
$$ $$
=
(\det \beta ) \sum_j \left[ (z-h-\bar \eta ) \left(  B L\right)_{ji}
 + \left( B\bar hBL \right)_{ji} \right] |\varphi_j >,
\eqno (4.36)
$$
where one recognizes the r.h.s.\ of (4.28).
The same generalization can be done for anti-\break
unitary transformations.
On the other hand, the invariance of a Slater determinant
under unitary transformations $ u $ of the orbitals,
with $\det u =1$, makes it possible to choose orbitals
$u \chi_i$ so that (4.26) transforms into (4.1).
The conditions (4.29) tell us that $L$ has to be a block matrix
like
$$
{\rm a)} \left(\matrix{l_1&0\cr
                    0&l_2\cr}\right)
\, {\rm or} \,\ \
   {\rm b)} \left(\matrix{0&l_1\cr
                    l_2&0\cr}\right),
\eqno (4.37)
$$
while $u$ is restricted to type a),
because the $\chi_i$ and $\chi_i^\prime $'s are only
partially antisymmetrized.
With $L$, $l_1$ and $l_2$ unitary there exists
a unitary matrix which diagonalizes them.
Choosing such a $ u$
brings (4.26) into
the form (4.1), with the $e^{i\theta_i}$ as eigenvalues.
For case a) one finds $\sigma (i) = i$ while in case b)
one has $\sigma^2 (i) = i$ where $\sigma (i)$ and $i$ refer to different
fragments.
In the next Section we use self consistent symmetries
to reduce the 16 equations
for the alpha-alpha system.
\bigskip

\centerline{5. The $\alpha$-$\alpha$ system}
\medskip
The elastic alpha-alpha collision has been studied by many authors [5]
as a test bench of microscopic theories of collisions.
As an application of antisymmetrized TIMF we consider elastic
${\alpha}-{\alpha}$ scattering in forward direction.
A central, spin- and isospin-independent interaction is chosen
as a sum of a short-range repulsion and a middle-range attraction,
both Gaussians
$$
 v(\vec{r},\vec{r'})=V_a \exp (-{{|\vec{r}-\vec{r'}|^2}\over{r_a^2}})
         +V_r \exp (-{{|\vec{r}-\vec{r'}|^2}\over{r_r^2}}) \,  .
\eqno{(5.1)}
$$
As potential depths $ V_a, V_r $ and ranges $ r_a,r_r $ we use the parameters
proposed by Volkov [6]
$$
V_a= -83.34 {\rm MeV} \, \, r_a= 1.6{\rm fm} \, \,
V_r= 144.86 {\rm MeV} \, \, r_r= 0.82{\rm fm} \, \, .
$$

The channel wavefunctions are described by boosted single nucleon Gaussians
$$
\chi(\vec{r},\vec{k}) = {\pi}^{-{ 3 \over 4}} {\beta}^{-{3\over 2}} \,
    \exp( i \vec{k} \cdot \vec{r} - {{r^2}\over{2{\beta}^2}} ) \, ,
\eqno{(5.2)}
$$
where $\vec{k}$ is the single particle boost.
Let $\vec{r}_1,\vec{r}_5$ denote the coordinates of the
proton with spin up in the projectile
and the target, respectively. In the same way, let $\vec{r}_2
,\vec{r}_6,\vec{r}_3,\vec{r}_7,\vec{r}_4,\vec{r}_8$
denote the protons spin down, the neutrons spin up and the neutrons spin
down, respectively.
The 8-particle channel wavefunction is therefore given by
$$
\chi(\vec{r}_1,...,\vec{r}_8,\vec{k})
= \chi_1(\vec{r}_1,\vec{k})\times...\times \chi_4(\vec{r}_4,\vec{k})\times
  \chi_5(\vec{r}_5,-\vec{k})\times...\times \chi_8(\vec{r}_8,-\vec{k})\,  .
\eqno{(5.3)}
$$

Our problem has the following symmetries:

a. Spin Isospin symmetries within each cluster imply

$$
\chi_1=\chi_2=\chi_3=\chi_4  \ \ {\rm and} \ \ \chi_5=\chi_6=\chi_7=\chi_8.
\eqno{(5.4)}
$$

b. Elastic scattering in forward direction, bra-ket symmetry:
$$
\chi_i(\vec{r}) \,  = \,  \chi_i'(\vec{r}) \ \ {\rm \,  for \ \, all \,\  i}
\eqno{(5.5)}
$$

c. Forward scattering together with time reversal:
$$
\chi_1^{\ast}(\vec{r})  =  \chi_5'(\vec{r}), \ \
\chi_5^{\ast}(\vec{r})  =  \chi_1'(\vec{r})
\eqno{(5.6)}
$$

d. Parity:
$$
\chi_1(\vec{r})  =  \chi_5(-\vec{r}), \ \
\chi_5^\prime (\vec{r})  =  \chi_1^\prime (-\vec{r}) .
\eqno{(5.7)}
$$

Following Section 4, we can choose trial functions which are consistent
with our variational equations
$$
\varphi_1(\vec{r})  =  \varphi_5(-\vec{r}), \ \
\varphi_5^\prime (\vec{r})  =  \varphi_1^\prime (-\vec{r})
\eqno{(5.8)}
$$ $$
\varphi_1^{\ast}(\vec{r})  =  \varphi_5^\prime (\vec{r}), \ \
\varphi_5^{\ast}(\vec{r})  =  \varphi_1^\prime (\vec{r}) \, .
\eqno{(5.9)}
$$
Spin and isospin symmetries are assumed again, naturally.
As described in Section 3, it is convenient to use a biorthogonal
representation :  If we rearrange the
trial functions into parity
eigenstates, such a representation can be
introduced as
$$
\tilde{\varphi}_+(\vec{r})  =
\sqrt{{1\over 2}}\, ( \varphi_1(\vec{r})+\varphi_5(\vec{r}) )
$$
$$
\tilde{\varphi}_-(\vec{r})  =
\sqrt{{1\over 2}} \, (-\varphi_1(\vec{r})+\varphi_5(\vec{r}))
$$ $$
\tilde{\varphi}_+'(\vec{r}) =
\sqrt{{1\over 2}}\, (\varphi_1'(\vec{r})+\varphi_5'(\vec{r}))
$$ $$
\tilde{\varphi}_-'(\vec{r})  =
\sqrt{{1\over 2}}\, (-\varphi_1'(\vec{r})+\varphi_5'(\vec{r}))
\,  .
\eqno{(5.10)}
$$
The symmetries and the spin-isospin independence of the chosen interaction
reduce the 16 coupled equations to only two coupled, non linear
equations for ${\tilde \varphi}_+$ and ${\tilde \varphi}_-$
which
have to be solved self consistently.

Because of our use of wave packets, zero point energies have to be taken
into account.
The zero point energy of a single nucleon is
$$
<\!\chi_i'|t_i|\chi_i\!>_{k=0} \,  = \,
E_0 \,  = \,  {3\over 4} {{\hbar^2}\over {m \beta^2}} \, ,
\eqno{(5.11)}
$$
where $m$ is the nucleon mass.   Note that $E_0$ is also the
center-of-mass zero point energy of an alpha particle.
The binding energy of the
alpha particle consists of the internal nucleon-nucleon
interaction and three zero point energies,
$$
E_{\alpha} \,  = \,  3E_0\,+\,6 <\!\chi_1' \chi_2'|v|\chi_1 \chi_2\!>
\,  .
\eqno{(5.12)}
$$
We adjust the width of the wave packet to 1.4 fm to get the experimental
binding energy of the alpha particle.
With wave packets the relative motion of the two alpha particles is
not a pure plane wave, but also a wave packet of Gaussian shape.
For that reason, a zero point correction for the relative motion
is necessary.

Because of the Galilean invariance of our many particle system,
the unphysical kinetic energy of the center-of-mass motion
has to be subtracted.
Hence the relative motion Hamiltonian reads
$$
H_r \,  = \,  H-T_{CM} \,  = \,  H- {1\over 16m}\sum_{i=1}^8 \vec{p_i}^2
\,  .
\eqno{(5.13)}
$$
The problem of $T_{CM}$ subtraction is the occurence of
non diagonal terms , such as $\vec{p}_1 \cdot \vec{p}_2$.
It is much simpler to approximate the center-of-mass
kinetic energy operator  $T_{CM}$
by its zero point expectation value,
$$
<\!T_{CM}\!> \,  = \,  E_0 \,  .
\eqno{(5.14)}
$$
With all these arguments, the total asymptotic energy consists of
the kinetic energy of the relative motion and the binding energy of the
two alpha particles as well as the zero point correction
of relative motion and the
center-of-mass subtraction
$$
E \,  = \,  {{\hbar^2}\over {4m}}(4k)^2+2E_{\alpha}+2E_0 \, .
\eqno{(5.15)}
$$
The two coupled equations for $\tilde{\varphi}_+$ and $\tilde{\varphi}_-$ are
solved numerically, using a scaled Hamiltonian and the channel wavefunction
as starting points. With the help of the Hamiltonian
$$
H_{\lambda} = \sum_i t_i+\lambda \sum_{i,j}{1 \over 2} v_{ij}
\eqno{(5.16)}
$$
we modify the stationarity conditions (2.2c) by introducing an additional
term
$$
\lambda {\cal A}V|\chi>+(\lambda-1)
\,{\cal A}\!\sum_{{i,j \epsilon 1, \cdots ,4}
\atop {i,j \epsilon 5, \cdots ,8}} \! {1 \over 2} v_{ij} |\chi> -
\, \, (z-H_{\lambda})|\Phi>\,=\,0 \, .
\eqno{(5.17)}
$$
We start with a large scaling parameter $\lambda$,
when Eq.(5.17) implies $\mid \Phi > \sim \mid \chi >$,
and reduce this parameter
to the physical value $\lambda=1$.
We also start with complex energies $E+i\Gamma$ and decrease $\Gamma$ to zero
to get the on-shell limit of $<\!\chi'| \Delta T|\chi\!>$.
A partial wave expansion for the mean field solution is useful to reduce the
numerical expense.
Choosing $\vec{k}$  along the z-axis, one can expand the channel
wave functions in partial waves with just magnetic quantum number $m=0$:
$$
r\chi_1(\vec{r})=\sum_l \chi_l(r)\,Y_l^0(\hat{r}),
\eqno{(5.18)}
$$
with
$$
\chi_l(r)=r\,\pi^{-{3\over 4}} \beta^{-{3\over 2}} \hat{l}
(4 \pi)^{1\over 2} i^l j_l(kr) \exp(-{{r^2}\over {2\beta^2}}) \, .
\eqno{(5.19)}
$$
In analogy we consider the following expansions of $\tilde{\varphi}_+$
and $\tilde{\varphi}_-$, see Eqs.(5.10),
$$
r \tilde{\varphi}_+(\vec{r})  =  \sum_{l \,even} \varphi_{l+}(r)\,
Y_l^0(\hat{r}) ,
$$ $$
r \tilde{\varphi}_-(\vec{r})  = \sum_{l \,odd} \varphi_{l-}(r)\,
Y_l^0(\hat{r})  ,
\eqno{(5.20)}
$$
where the summation can be restricted to even and odd numbers, respectively.
E.g., the left hand side of our mean field Equation (3.12a)
in partial wave expansion
for $\tilde{\varphi}_+$ then reads
$$
\left[ \eta_+ + {\hbar^2 \over 2m}{{d^2}\over{dr^2}} -
{\hbar^2{l(l+1)}\over{2mr^2}} \right]\ \varphi_{l+}(r)
- 3\sum_{l'} U_{ll'}^{++}(r) \varphi_{l'+}(r)
- 4\sum_{l'} U_{ll'}^{--}(r) \varphi_{l'+}(r)
$$ $$
+ \sum_{l'} U_{ll'}^{-+}(r) \varphi_{l'-}(r),
\eqno{(5.21)}
$$
where
$$
U_{ll'}^{\pi\nu}(r)  =  \sum_{L} \hat{L} \hat{l} \hat{l'}
 \left( {l\atop 0}{l'\atop 0}{L\atop 0} \right)^2 U_L^{\pi\nu}(r) ,
\eqno{(5.22)}
$$ $$
U_L^{\pi\nu}(r)  =  \int\!dr' C_L(r,r') \sum_j \sum_{j'}
\hat{j} \hat{j'} \hat{L} \left( {j\atop 0}{j'\atop 0}{L\atop 0} \right)^2
$$ $$
\times \cases{  \varphi_{j+}(r')\varphi_{j'+}(r')/n_{++}, &
                  for $(\pi\nu) = (++)$; \cr
                \varphi_{j-}(r')\varphi_{j'-}(r')/n_{--}, &
                  for $(\pi\nu) = (--)$; \cr
                \varphi_{j-}(r')\varphi_{j'+}(r')/n_{--}, &
                  for $(\pi\nu) = (-+)$, \cr}
\eqno{(5.23)}
$$ $$
C_L(r,r')   =   V_a\,I_L({{2 r r'}\over{r_a^2}}) \exp (-{{r^2+{r'}^2}
\over {r_a^2}})
 +V_r\,I_L({{2 r r'}\over{r_r^2}}) \exp (-{{r^2+{r'}^2}\over{r_r^2}}),
\eqno{(5.24)}
$$
$$
n_{++}   =   <\!\tilde{\varphi}_+'|\tilde{\varphi}_+\!> ,\
n_{--}   =   <\!\tilde{\varphi}_-'|\tilde{\varphi}_-\!> ,
\eqno{(5.25)}
$$
and $\eta_+$ is the self energy corresponding to $\tilde{\varphi}_+$.
Note that in general $\eta_-$  is different from $\eta_+$
because of the chosen representation.
Here $I_L$ denotes the modified spherical Bessel function of order L.
Also note that in Eq.(5.21), when written with explicit parity
coupling, only local potential terms appear.

Corresponding to the expansion of the l.h.s.\ of Eq.(3.12a)
an expansion of the source term is possible by using $\chi$
instead of $\tilde{\varphi}$. A channel potential
$S_{ll'}^{+,1}$ acting on $\chi_{l'1}$ can be defined as
$$
S_{ll'}^{+,1}(r) =
 \sum_{L} \hat{L} \hat{l} \hat{l'}
 \left( {l\atop 0}{l'\atop 0}{L\atop 0} \right)^2 S_L^{+,1}(r) ,
\eqno{(5.26)}
$$
with
$$
S_L^{+,1}(r) =
\int\!dr' C_L(r,r') \sum_j \sum_{j'}
\hat{j} \hat{j'} \hat{L} \left( {j\atop 0}{j'\atop 0}{L\atop 0} \right)^2
\varphi_{j+}(r') \chi_{j'1}(r').
\eqno{(5.27)}
$$

The spin and isospin symmetry reduces the
large amount of cofactors of first, second and third order
to only cofactors between nucleons with same spin and isospin.
The convergence of our method is usually fast.
On the way to the on-shell limit
neither bifurcation nor crossing occurred. As described in [7],
solutions with incoming waves sometimes
creep in. Standard relaxation techniques [8]
change these solutions into outgoing ones. Relaxation techniques are also
useful to speed up the convergence of our iterative method and remove
other unphysical solutions.

The Born terms of both the antisymmetrized ($\diamond$)
and the non-antisymmetrized ($\circ $)
theories,  respectively, are shown in Figure 1.
Note their different behaviours  when the boost $k$ diminishes.
This is an obvious illustration of the importance of exchange terms
at low energies, whereas the Pauli principle may be
neglected at high energies.
One should, however, realize that the wave packets
$\chi$, $\chi^\prime$ whose
widths are related to the sizes of the fragments,
may poorly correspond,
at low energy, to well separated asymptotic
fragments as required when using
the above formula for the antisymmetrized transition amplitude.
In consequence, the antisymmetrized wavepacket
${\cal A} |\chi >$ has vanishing
norm as $k$ tends to zero, and the antisymmetrization effect seen in
$<\chi^\prime|V{\cal A}|\chi >$
is likely to be too strong at low $k$-values:
If the relative motion part of $\chi$ were a plane wave,
one would only have to
exclude $k=0$ where the fragments fully overlap;
since we deal with
wave packets, we have to exercise care in a finite region around $k=0$.
This defect can be remedied completely by
taking suitable superpositions [9]
at different $k$-values on both the
bra and the ket sides in the transition
amplitude. Rather than going into such an
elaborate calculation we restrict
ourselves to estimating the correction by dividing
$< \chi^\prime |V {\cal A} | \chi >$ by the
required\footnote*{Note that $V$ and $\cal A$ do not commute,
$< \chi^\prime |V {\cal A} | \chi > \ne
 < \chi^\prime |{\cal A}V {\cal A} | \chi >$,
and renormalization is not required on the bra side.
}
normalization factor
$\sqrt{ < \chi | {\cal A} | \chi > } $ on the ket side.
As a result the antisymmetrized Born term is modified at low $k$-values
as shown by the ($\triangle $) curve in Fig.1. The limiting value remains
zero nevertheless.

Figure 2 displays the real parts of $<\!\chi'|\Delta T|\chi\!>$
as a function of the total asymptotic energy E for both the antisymmetrized
and the non antisymmetrized calculations.
In the same way, Figure 3 shows the energy dependence of the relative
difference $\Delta$ between the antisymmetrized
and non antisymmetrized results
for the imaginary part of the multistep amplitude
$$
\Delta = {{ <\!\chi'|{\rm Im} \Delta T|\chi\!>_{NA}
- <\!\chi'|{\rm Im} \Delta T|\chi\!>_{AS} }\over
{<\!\chi'|{\rm Im}\Delta T|\chi\!>_{NA}}}
\, .
\eqno{(5.28)}
$$
As expected, antisymmetrization is important in the low-energy region.
Regarding the defect of the channel waves $\chi$, $\chi^\prime$ at low
energies, the same arguments apply as given above for the Born term.

Finally we show on Fig.4 an example of one of the wave functions $\varphi$
generated by our code. It is remarkable that it is a square integrable state,
although we are dealing with a theory of collisions. The reason for this is
not only that the source terms $\chi , \chi^\prime$ are wave packets. Rather,
a main feature of this formalism is the ocurrence of retarded self energies
 $\eta$ with {\it finite} imaginary parts even when the imaginary part
 of the many body energy $E$ vanishes on shell.
We notice on Fig.4 the damped
oscillations of $\varphi_-$
away from the center
of the interaction region.
 The damping is indeed due to the retardation expressed
by Im $\eta$.
\bigskip
\centerline{6. Discussion and Conclusion}
\medskip
At low energies not exceeding a few MeV per nucleon, previous
 microscopic theories of collisions [5] have been quite successful
 in the description of nuclear collisions. This success can be understood,
to a large extent at least, from the limited number of channels open to the
reaction mechanisms, as analyzed, e.g. by Tang [10]. Conversely, at large
 energies, beyond those corresponding to the Fermi momentum,
 effective theories have been also successfully proposed [11]
for the derivation of
optical potentials between the colliding nuclei. There is still room for a
new microscopic approach, as shown by the present paper.

 Indeed we have proved here
that the TIMF antisymmetrized theory can be implemented
numerically, although it
seems to be somewhat demanding and ambitious
in activating {\it all} degrees of
freedom. The TIMF equations involve all the single particle wave
functions on the same footing,
while we can insist that low energy [5] as well
 as high energy [11] theories essentially
privilege the wave functions and/or the interactions governing the relative
 degrees of freedom between the clusters. These TIMF equations
 generate {\it non perturbatively} intermediate states $\Phi , \Phi^\prime$
which parametrize the multistep transition amplitude and give a picture of
the reaction mechanism, as illustrated by Fig.4.

Once again we want to stress
the fact that we are using a wave packet
representation of the T-matrix rather than the more physical plane wave
representation of relative motion.
 The conversion of
one representation into another [9] is a typical generator coordinate [12]
 problem, which we intend to study.

Finally, it is clear that the mean field calculation which has been presented
 here can accomodate higher order corrections. An advantage of the TIMF
technique is its close connection with the traditional techniques of the shell
model. Therefore Tamm-Dancoff corrections
and even RPA corrections are available.

Some parts of this paper are also part of the Diplomarbeit of A. Wierling
and the
Doktorarbeit of J. Lemm at the University of M\"unster, respectively.
This work has benefited from a Procope grant for franco-german
scientific cooperation.
Two of us (B.G. and A.W.) thank the nuclear theory
group of Kyoto University, where part of this work was done,
for hospitality.
\bigskip
\centerline{\bf References}
\smallskip
1. Austern, N.: Direct Nuclear Reaction Theory. Wiley, New York, 1970

2. Giraud, B.G., Nagarajan, M.A., Thompson, I.J.:
: Annals of Phys. 152, 475 (1984)

3. Bonche, P., Koonin, S., Negele, J.: Phys. Rev. C13, 1226 (1976)

4. Giraud, B.G., Nagarajan, M.A., Noble, C.J.,: Phys. Rev. A 34, 1034 (1986)

5. Schmid, E.W., Wildermuth, K.: Nucl. Phys. 26, 463 (1961)

$\,\,$   Tanabe, F., Tohsaki, A., Tamagaki, R.,:
Prog. Theor. Phys. 53, 671 (1975)

$\,\,$   Lumbroso, A.,: Phys. Rev C10, 1271 (1974)

$\,\,$   Friedrich, H.: Phys. Lett. C 74, 211 (1981)

6. Volkov, A.B.,: Nucl. Phys. 74, 33 (1965)

7. Giraud, B.G., Kessal, S., Weiguny, A.: J. Math. Phys. 29, 2084 (1988)

8. Stoer, J., Bulirsch, R.: Introduction to Numerical Analysis.
Springer,

$\,\,$ New York, 1983

9. Giraud, B.G., Lemm, J., Weiguny, A.,
  Wierling, A.: Z.Phys. A 343, 249
 (1992)

10. Tang, Y.C.: Kyoto Symposium on Nuclear Clustering and Related
Problems,

$\,\,$   Aug.31-Sept.2, 1992,
Yukawa Insitute, Kyoto, Japan (Y. Fujiwara ed.)

11. Wada, T., Horiuchi, H.: in Proceedings of the
International Symposium on

$\,\,$ Developments of Cluster Dynamics,
Aug.1-3, 1988, (Eds. Akaishi, Y.,
Kato,

$\,\,$  K., Noto, H., Okabe, S.) p136, and references therein.

$\,\,$ Ohtsuka, N., Linden, R., Faessler, A., Malik, F.B.:
Nucl. Phys. A 465, 550 (1987)

$\,\,$ Michel, F., Vanderpoorten, R.: Phys. Lett. 82B, 183 (1979)

12. Horiuchi, H.: Prog. Theor. Phys. 43, 375 (1970)

13. Rheinhardt, H.: Nucl.Phys. A390, 70 (1982)

\bigskip
\bigskip
\bigskip
\bigskip
\centerline{Appendix \ \  Channel Spin}
\medskip
The prior and post interactions $V$, $V^\prime$ are symmetric
under nucleon exchange {\it within} the respective fragments but
unsymmetric when nucleon exchange {\it between} fragments is included.
Under this partial lack of permutation symmetry, only pure product or
intrafragment antisymmetrized wave
functions $\chi$, $\chi^\prime$
were used so far in this paper.
This is similar to the second quantized
formulation of TDMF by Reinhardt [13].
We shall now show that there exists an
equivalent "channel spin" formulation
which works with totally antisymmetrized,
two-component channel wave functions.
For the sake of clarity of presentation,
we shall slightly modify our notation
of Sections 2 and 3 in the following way:
We denote the pure products of (normalized)
single-particle wave functions
in initial and final channels, eq(3.1a),
by round brackets $ |\chi )$
and $ (\chi^\prime |$, respectively.
Curly brackets are used for channel wave functions
$$ | \chi \} = \sqrt{N_a ! N_b !}
\, {\cal A }_a {\cal A }_b | \chi )
    \; , \eqno (A.1a) $$
 $$ \{ \chi^\prime | = \sqrt{N_c ! N_d !} \, ( \chi^\prime |
 {\cal A }_c {\cal A }_d
    \; , \eqno (A.1b)  $$
which are normalized and antisymmetrized
{\it within} the fragments
$a$, $b$, $c$, $d$, in obvious notation.
Angular brackets, like in $|\Phi >$,
are reserved for fully antisymmetrized wave functions,
$$
|\chi> = {1 \over {\sqrt N !}} \sum_{\cal P} (-)^{\cal P} {\cal P} |\chi).
\eqno (A.1c)
$$

With such  $|\chi \}$, $\{ \chi^\prime |$ it is in general
not possible to diagonalize all three overlap matrices
$\alpha_{ij} = < \varphi_i^\prime | \chi_j > $,
$\alpha^\prime_{ij} = < \chi_i^\prime | \varphi_j > $,
$\beta_{ij} = < \varphi_i^\prime | \varphi_j > $
by rearrangement of orbitals within the four sets of wave functions
$\{ \chi_i \} $, $\{ \chi^\prime_i \}$,
$\{ \varphi_i \} $, $\{ \varphi^\prime_i \}$.
This would correspond to solve the following set of equations
for the matrices $M $, $M^\prime $  and $B $, $B^\prime $
where the latter two have block structures which do not mix
orbitals of different fragments,
 $$ M^\prime \beta M = d_\beta \; ,$$
 $$ M^\prime \alpha B = d_\alpha \; , \eqno (A.2) $$
 $$ B^\prime \alpha^\prime M = d_{\alpha^\prime} \; ,$$
with diagonal matrices $d_\alpha $, $d_{\alpha^\prime} $,
$d_\beta $.
This means that $B^{-1} \alpha^{-1} \beta \alpha^{\prime -1}
B^{\prime -1} = d_\alpha^{-1} d_\beta d_{\alpha^\prime}^{-1} $
has to be diagonal, which is not possible in general
because of the block structure of $B $ and $B^\prime $,
which is a consequence of partial antisymmetrization.
To diagonalize all three overlap matrices one has to
work with totally antisymmetrized channel wave functions
$\chi $ and $\chi^\prime $.

Because the
actions of $V$ and $(H-E)$
on an exact channel eigenstate  are equivalent,
one could replace
 $V$ by $ ( H - E ) $ which commutes with the total
antisymmetrizer $\cal A $.
However, the equivalence of $(H-E)$ and $V$
(or $V^\prime $) is destroyed
if approximate channel wave functions $\chi$, $\chi^\prime$ are used,
like in the mean field approximation, when $(H-E)|\chi >$ and
$V|\chi >$ may differ by long-ranged terms.
Because $V$ is normally short-ranged, $V$ should be used.
In order to alleviate the clumsy partial
antisymmetrization imposed by
$V$ and $V^\prime$, the concept of channel (pseudo-)spin
is introduced in this Appendix.

Inside the matrix elements which involve $\chi $ or $\chi^\prime $,
the single-particle wave functions $\chi_i $, $\chi^\prime_i $
and $\varphi_i $, $\varphi^\prime_i $
are translated into two-component
(channel spin) wave functions
 $$ | \chi_i > \rightarrow
    | \chi^{cs}_i > =
    \left| \matrix{\chi_i \cr 0\cr }\right>  ,
     \qquad {\rm for } \; i \, \epsilon \, a ,
 \eqno (A.3a) $$
 $$  |\chi_i > \rightarrow
     | \chi_i^{cs} > =
    \left| \matrix{0 \cr \chi_i \cr }\right> ,
     \qquad {\rm for } \; i \,\epsilon \, b ,
    \eqno (A.3b)
     $$
 $$ < \chi_i^\prime | \rightarrow
    < \chi^{\prime cs}_i | =
    \left< \matrix{\chi_i^\prime \cr 0 \cr }\right| ,
     \qquad {\rm for } \; i \, \epsilon \, c ,
 \eqno (A.3c) $$
 $$ < \chi_i^\prime | \rightarrow
    < \chi^{\prime cs}_i | =
    \left< \matrix{0 \cr \chi_i^\prime \cr }\right| ,
     \qquad {\rm for } \; i \, \epsilon \, d ,
 \eqno (A.3d) $$
 $$  |\varphi_i > \rightarrow
     | \varphi_i^{cs} > =
    \left| \matrix{\varphi_i \cr \varphi_i \cr }\right> ,
    \eqno (A.3e)
     $$
 $$ < \varphi_i^\prime | \rightarrow
    < \varphi^{\prime cs}_i | =
    \left< \matrix{\varphi_i^\prime \cr \varphi_i^\prime \cr }\right| .
 \eqno (A.3f) $$
 Therefore we consider the transformation
 $$ | \chi \} \rightarrow
    | \chi^{cs} \} =
    \left(
    \sqrt{N_a !} {\cal A}_a \prod_{i \epsilon a}
    \left| \matrix{\chi_i \cr 0\cr }\right>
    \right)
    \left(
    \sqrt{N_b !} {\cal A}_b \prod_{j \epsilon b}
    \left| \matrix{0 \cr \chi_j \cr }\right>
    \right) ,
    \eqno (A.4a)
     $$
 $$ \{ \chi^\prime | \rightarrow
     \{ \chi^{\prime cs} | =
    \left(
    \sqrt{N_c !}
    \prod_{i \epsilon c}
    \left< \matrix{\chi_i^\prime \cr 0 \cr }\right|
     {\cal A}_c
    \right)
    \left(
    \sqrt{N_d !}
    \prod_{j \epsilon d}
    \left< \matrix{0 \cr \chi_j^\prime \cr }\right|
     {\cal A}_d
    \right)  .
    \eqno (A.4b)
    $$
For $\chi$ a channel spin 'up' means that the orbital
belongs to fragment $a$, and channel spin 'down' means that
it belongs to fragment $b$. For $\chi^\prime$ the distinction
is between fragments $c$ and $d$.
Equal components in channel spin formalism, eqs. (A.3e) and
(A.3f), are assigned to single-particle orbitals
$\varphi_i$, $\varphi^\prime_i$ of
trial functions $\Phi$, $\Phi^\prime$.
This assignment has no effect on total antisymmetrization,
 $$ | \Phi > \rightarrow
    | \Phi^{cs} > =
    \sqrt{N !} {\cal A} \prod_i
    \left| \matrix{\varphi_i \cr \varphi_i \cr } \right> ,
\eqno (A.4c)
    $$

 $$ < \Phi^\prime | \rightarrow
     < \Phi^{\prime cs} | =
    \sqrt{N !}
    \prod_{i}
    \left< \matrix{\varphi_i^\prime \cr \varphi_i^\prime \cr } \right|
     {\cal A}  .
\eqno (A.4d)
    $$

 Now we introduce channel spin dependent potentials
 which replace $V$ and $V^\prime $
 but leave their matrix elements invariant.
 With the Pauli matrix $\sigma_z^i$
 acting on particle $i$ in
 $\chi $-channel spin representation
 and
 with $\sigma_z^{\prime i} $ acting on particle $i$ in
 $\chi^\prime $-channel spin representation,
the replacement of the interaction $v_{ij} $
between particle $i$ and $j$ by
$$ v^{cs}_{ij} =  {1\over 2} (1-\sigma_z^i \sigma_z^j ) v_{ij}
\; \; \hbox{\rm in $\chi$-channel spin representation} ,
\eqno (A.5a) $$
$$ v^{\prime cs}_{ij} =  {1\over 2}
(1-\sigma_z^{\prime i} \sigma_z^{\prime j} ) v_{ij}
\; \; \hbox{\rm in $\chi^\prime $-channel spin representation} ,
\eqno (A.5b) $$
 sets  the interactions $v_{ij}^{cs}$, $v_{ij}^{\prime cs}$
to zero if acting within fragments
and leaves them equal to $v_{ij}$
 otherwise.
With the definition
 $$ V=\sum_{{i \epsilon a}\atop{j \epsilon b}} v_{ij} \rightarrow
 V^{cs} = \sum_{i,j} {1\over 2} (1-\sigma_z^i \sigma_z^j ) v_{ij}  ,
\eqno (A.5c) $$
 $$ V^\prime =\sum_{{i \epsilon c}\atop{j \epsilon d}} v_{ij}
 \rightarrow
 V^{\prime cs}=
 \sum_{i,j} {1\over 2}
 (1-\sigma_z^{\prime i} \sigma_z^{\prime j} ) v_{ij} ,
\eqno (A.5d) $$
 it is easily seen that
$$<\Phi^\prime | V | \chi \} =
<\Phi^{\prime cs} | V^{cs} | \chi^{cs} \} ,
\eqno (A.6a) $$
$$\{ \chi^\prime | V^\prime | \Phi > =
\{ \chi^{\prime cs} | V^{\prime cs} | \Phi^{cs} > .
\eqno (A.6b) $$
In this enlarged channel spin space $V^{cs}$ and $V^{\prime cs}$
commute with $\cal A$,
$$ [ V^{cs} , {\cal A} ] = 0 =
 [ V^{\prime cs} , {\cal A} ] ,
\eqno (A.7) $$
and therefore
 $$ <\Phi^{\prime cs} | V^{cs} | \chi^{cs} \}
= <\Phi^{\prime cs} | {\cal A} V^{cs} | \chi^{cs} \} =
$$
$$
 \sqrt{N_a ! N_b !}
<\Phi^{\prime cs} | V^{cs}
{\cal A}
{\cal A}_a
{\cal A}_b
| \chi^{cs} )
=\sqrt{{N_a ! N_b ! }\over{N !}}
 <\Phi^{\prime cs} | V^{cs} | \chi^{cs} > ,
\eqno (A.8a) $$
and correspondingly with (A.1b)
 $$ \{ \chi^{\prime cs} | V^{\prime cs} | \Phi^{cs} >
=\sqrt{{N_c ! N_d ! }\over{N !}}
  <\chi^{\prime cs} | V^{\prime cs} | \Phi^{cs} > .
\eqno (A.8b) $$
We recall here that ${\cal A A}_a = {\cal A A}_b = {\cal A}$.
It is thus possible
to work with fully antisymmetrized wave functions only.
The functional $F$ of Eq.(2.5c)
then reads in the channel spin formulation
 $$ F=
     \sqrt{{N_a ! N_b ! }\over{N !}}
     <\Phi^{\prime cs} | V^{cs} | \chi^{cs} >
   + \sqrt{{N_c ! N_d ! }\over{N !}}
     <\chi^{\prime cs} | V^{\prime cs} | \Phi^{cs} >
   -<\Phi^{\prime} | ( z - H ) | \Phi > ,
\eqno (A.9)
$$
  where the channel spin is not needed for the
  $<\Phi^{\prime} | (z - H) | \Phi >$ matrix element.
  (An introduction of channel spin in this matrix element
is nevertheless possible and would require a normalization factor
$1\over \sqrt{2} $ for every orbital
$\varphi_i$, $\varphi_i^\prime$,
but has no further effect.
Note that this normalization factor must be included inside
  $<\Phi^{\prime} | (z - H) | \Phi >$ only.)

  Within $\chi^{cs}$ and $\chi^{\prime cs}$ the orbitals can now be rearranged
  like those within $\Phi$ and $\Phi^{\prime}$.
  This leads to generalized channel spin orbitals
 $$
  |\chi_i^{cs} > =
  \left| \matrix{\chi_i^1 \cr \chi_i^2 \cr} \right> ,
\eqno (A.10a) $$
 $$ <\chi_i^{\prime cs} | =
  \left< \matrix{\chi_i^{\prime 1}
  \cr \chi_i^{\prime 2} \cr} \right| .
\eqno (A.10b) $$
It is now possible to diagonalize the generalized
overlap matrices
$<\varphi_i^{\prime cs} | \chi_j^{cs} >$,
$<\chi_i^{\prime cs} | \varphi_j^{cs} >$
and corresponding generalized cofactors
$M^{cs}_{ij}$,
$M^{\prime cs}_{ij}$ through linear rearrangement of
the $\chi_i^{cs}$, $\chi_i^{\prime cs}$ leaving the
$\varphi_i$,
$\varphi_i^\prime $ unchanged.
This enables us
to diagonalize
$<\varphi^\prime_i | \varphi_j >$
and $<\varphi^\prime_i | h | \varphi_j >$ by
linear rearrangement of the
$\varphi_i$,
$\varphi_i^\prime $.
In Eqs.(3.4a,b) the sum ${1\over 2} \sum_{\alpha \beta}  $
has to be replaced by ${1\over 4} \sum_{k l}  $ where the indices
$k$, $l$ run over all orbitals because of the totally
antisymmetrized wave functions
$$
< \Phi^{\prime cs}| V^{cs} | \chi^{cs} > =
{1\over 4}
\sum_{i j k l}
< \varphi_i^{\prime cs} \varphi_j^{\prime cs}
   | v^{cs} | \chi_k^{cs} \chi_l^{cs} >
M_{ijlk}^{cs}  ,
\eqno (A.11a)
$$
$$
< \chi^{\prime cs} | V^{\prime cs} | \Phi^{cs} > =
{1\over 4}
\sum_{i j k l}
< \chi_k^{\prime cs} \chi_l^{\prime cs}
   | v^{\prime cs} | \varphi_i^{cs} \varphi_j^{cs} >
M_{klji}^{\prime cs}    .
\eqno (A.11b)
$$
Here antisymmetrized matrix elements are used for $v^{cs}$
and $v^{\prime cs}$.

 The derivatives with respect to the single-particle orbitals
 $\varphi_i$, $\varphi_i^\prime$
must take into account the fact that $\varphi_i$ resp.
$\varphi_i^{\prime}$ appear twice inside the channel spin
trial functions.
With the relations for generalized channel spin wave functions
$$  <\varphi_m^{\prime cs} \varphi_n^{\prime cs}
    | v^{cs} | \chi_k^{cs} \chi_l^{cs} >
  = <\varphi_m^\prime \varphi_n^\prime
    | v | \chi_k^1 \chi_l^2 + \chi_k^2 \chi_l^1 >    ,
\eqno (A.12a)
$$
$$  <\chi_m^{\prime cs} \chi_n^{\prime cs}
    | v^{\prime cs} | \varphi_k^{cs} \varphi_l^{cs} >
  = <\chi_m^{\prime 1} \chi_n^{\prime 2}
  + \chi_m^{\prime 2} \chi_n^{\prime 1}
    | v | \varphi_k \varphi_l >
\eqno (A.12b)
$$
for the antisymmetrized matrix elements, Eqs.(3.5a,b)
become:
$$
\eqalignno{
{{\delta C}\over {\delta <\varphi_i^{\prime} |}}
&
=\sqrt{{N_a ! N_b ! }\over{N !}}
\Biggl(
     {1\over 2} \sum_{jkl}
      < \cdot \, \varphi_j^{\prime} | v |
              \chi_k^{1} \chi_l^2 + \chi_k^2 \chi_l^1 >
        M_{ijlk}^{cs}
& (A.13a)
\cr
&\qquad
      +{1\over 4} \sum_{jmnkl}
        < \varphi_m^\prime \varphi_n^{\prime} | v |
                    \chi_k^{1} \chi_l^2 + \chi_k^2 \chi_l^1 >
         M_{mnijlk}^{cs}
         |\chi_j^1 + \chi_j^2 >
 \Biggr),
\cr }
$$
$$
\eqalignno{
{{\delta C^\prime }\over {\delta |\varphi_i > }}
&
=\sqrt{{N_c ! N_d ! }\over{N !}}
\Biggl(
     {1\over 2} \sum_{jkl}
      < \chi_k^{\prime 1} \chi_l^{\prime 2}
      + \chi_k^{\prime 2} \chi_l^{\prime 1}
      | v |
              \cdot \, \varphi_j >
        M_{klji}^{\prime cs}
& (A.13b)
\cr & \qquad
      +{1\over 4} \sum_{jmnkl}
        < \chi_k^{\prime 1} \chi_l^{\prime 2}
        + \chi_k^{\prime 2} \chi_l^{\prime 1}
         | v |
                    \varphi_m \varphi_n >
         M_{kljinm}^{\prime cs}
        <\chi_j^{\prime 1} + \chi_j^{\prime 2} |
 \Biggr) .
\cr }
 $$
Now the self consistent channel potentials can be defined by
$$
\eqalignno{
 < \cdot \, | S_1 | \chi_l^1 >
&
 = {1 \over {<\Phi^{\prime cs} | \chi^{cs} >}}
   \sum_{n k}
   < \cdot \, \varphi_n^\prime  | v | \chi_l^1 \chi_k^2 >
  M_{nk}^{cs},
& (A.14a)
\cr
 < \cdot \, | S_2 | \chi_l^2 >
&
 = {1 \over {<\Phi^{\prime cs} | \chi^{cs} >}}
   \sum_{n k}
   < \cdot \, \varphi_n^\prime | v | \chi_l^2 \chi_k^1 >
   M_{nk}^{cs} ,
& (A.14b)
\cr
 < \chi_l^{\prime 1}  | S_1^{\prime } | \, \cdot >
&
 = {1 \over {<\chi^{\prime cs} | \Phi^{cs} >}}
   \sum_{k n}
   < \chi_l^{\prime 1} \chi_k^{\prime 2} | v |
    \cdot \, \varphi_n >
   M_{k n}^{\prime cs} ,
& (A.14c)
\cr
 < \chi_l^{\prime 2}  | S_2^{\prime } | \, \cdot >
&
 = {1 \over {<\chi^{\prime cs} | \Phi^{cs} >}}
   \sum_{k n}
   < \chi_l^{\prime 2} \chi_k^{\prime 1} | v |
    \cdot \, \varphi_n >
   M_{kn}^{\prime cs} .
& (A.14d) \cr
}
 $$
 Then Eqs.(3.5c,d) and (3.6e,f) do not change
 because there are no $\chi$, $\chi^\prime$ involved,
  but Eqs.(3.7a,b)
 now read
 $$
\eqalignno{
{{\delta C}\over {\delta <\varphi_i^{\prime} |}}
&
=\sqrt{{N_a ! N_b ! }\over{N !}}
<\Phi^{\prime cs} | \chi^{cs} >
\Biggl[
 \sum_j A_{ji}^{cs}
 \bigl( S_1 | \chi_j^1 > + S_2 | \chi_j^2 > \bigr)
 \cr & \qquad
 +{1\over 2}
  \sum_{jml}
  \left(
           < \varphi_m^\prime | S_1 | \chi_l^1 >
          +< \varphi_m^\prime | S_2 | \chi_l^2 >
     \right) A_{ji}^{cs} A_{lm}^{cs} |\chi_j^1 + \chi_j^2 >
\cr & \qquad
  -\sum_{jml} \left( <\varphi_m^\prime | S_1 | \chi_l^1 >
            + <\varphi_m^\prime | S_2 | \chi_l^2 > \right)
             A_{jm}^{cs} A_{li}^{cs} |\chi_j^1 + \chi_j^2 >
\Biggr] ,
& (A.15a)
\cr }
$$
 $$
\eqalignno{
{{\delta C^\prime}\over {\delta |\varphi_i >}}
&
=\sqrt{{N_c ! N_d ! }\over{N !}}
<\chi^{\prime cs} | \Phi^{cs} >
\Biggl[
 \sum_j A_{ij}^{\prime cs} ( <\chi_j^{\prime 1} | S_1^\prime
 + < \chi_j^{\prime 2} | S_2^\prime )
 \cr & \qquad
   + {1\over 2}
    \sum_{jml}
      \left(
           < \chi_l^{\prime 1} | S_1^\prime | \varphi_m >
          +< \chi_l^{\prime 2} | S_2^\prime | \varphi_m >
     \right)
     A_{ij}^{\prime cs} A_{ml}^{\prime cs} <\chi_j^{\prime 1}
     + \chi_j^{\prime 2} |
\cr & \qquad
  - \sum_{jml}
             \left(
           < \chi_l^{\prime 1} | S_1^\prime | \varphi_m >
          +< \chi_l^{\prime 2} | S_2^\prime | \varphi_m >
     \right)
     A_{mj}^{\prime cs} A_{il}^{\prime cs} <\chi_j^{\prime 1}
     + \chi_j^{\prime 2} |
\Biggr] .
& (A.15b)
\cr }
$$
These equations are invariant under linear
rearrangement of the orbitals $\chi_i^{cs} $, $\chi_i^{\prime cs}$,
and they reduce to Eqs.(3.7a,b) when the wave functions (A.3a,b,c,d)
are used.
In a representation where both overlap matrices
 $<\varphi_i^{\prime cs} | \chi_j^{cs} >$
 and
 $<\chi_i^{\prime cs} | \varphi_j^{cs} >$
 are diagonal, equations (A.15a,b) reduce to
$$
\eqalignno{
{{\delta C}\over {\delta <\varphi_i^{\prime} |}}
&
=\sqrt{{N_a ! N_b ! }\over{N !}}
{<\Phi^{\prime cs} | \chi^{cs} >
\over
<\varphi_i^{\prime } | \chi_i^{1} +\chi_i^{2} >}
\Biggl[
  \bigl( S_1 | \chi_i^1 > + S_2 | \chi_i^2 > \bigr)
 \cr & \qquad
 + {1\over 2}
          \Biggl(
 \sum_{l}
          {< \varphi_l^\prime | S_1 | \chi_l^1 >
          +< \varphi_l^\prime | S_2 | \chi_l^2 >
          \over
          <\varphi_l^{\prime } | \chi_l^{1} +\chi_l^{2} >}
          \Biggr)
         |\chi_i^1 + \chi_i^2 >
\cr & \qquad
  -\sum_{j}
          {
          <\varphi_j^\prime | S_1 | \chi_i^1 >
            + <\varphi_j^\prime | S_2 | \chi_i^2 >
          \over
          <\varphi_j^{\prime } | \chi_j^{1} +\chi_j^{2} >}
              |\chi_j^1 + \chi_j^2 >
\Biggr] ,
& (A.16a)
\cr }
$$
$$
\eqalignno{
{{\delta C^\prime }\over {\delta |\varphi_i >}}
&
=\sqrt{{N_c ! N_d ! }\over{N !}}
{<\chi^{\prime cs} | \Phi^{cs} >
\over
<\chi_i^{\prime 1} +\chi_i^{\prime 2} | \varphi_i >}
\Biggl[
  \bigl( < \chi_i^{\prime 1} | S_1^\prime
  + < \chi_i^{\prime 2} | S_2^\prime \bigr)
 \cr & \qquad
 + {1\over 2}
          \Biggl(
 \sum_{l}
          {< \chi_l^{\prime 1} | S_1^\prime | \varphi_l >
          +< \chi_l^{\prime 2} | S_2^\prime | \varphi_l >
          \over
          <\chi_l^{\prime 1} +\chi_l^{\prime 2} | \varphi_l >}
          \Biggr)
         <\chi_i^{\prime 1} + \chi_i^{\prime 2} |
\cr & \qquad
  -\sum_{j}
       {
       <\chi_i^{\prime 1}| S_1^\prime | \varphi_j >
         + <\chi_i^{\prime 2} | S_2^\prime | \varphi_j >
        \over
       <\chi_j^{\prime 1} + \chi_j^{\prime 2} | \varphi_j >}
           <\chi_j^{\prime 1} + \chi_j^{\prime 2} |
\Biggr] .
& (A.16b)
\cr }
$$
 Note that
$$
 <\Phi^{\prime cs} | \chi^{cs} >
 = \prod_i <\varphi_i^\prime | \chi_i^1 +\chi_i^2 >
\, {\rm and} \,
 <\chi^{\prime cs} | \Phi^{cs} >
 = \prod_i <\chi_i^{\prime 1} + \chi_i^{\prime 2} | \varphi_i > .
\eqno (A.17)$$
 These equations can now
be combined with Eqs.(3.8c,d) where
$<\varphi^\prime_i | \varphi_j >$
and $<\varphi^\prime_i | h | \varphi_j >$ are diagonal.
Because all the cofactors are simultaneously diagonalized
when the three overlap matrices are chosen diagonal, the enormous
number of cofactors, like $N^6$ for $M_{mnijlk}$ with six indices,
is reduced to the number of diagonal elements only.
The price to be paid for this simplification is
that one has to deal with two component wave functions.
They give rise to additional terms in the inhomogenous parts
of (A.16a,b) as compared to (3.7a,b).
A generalization of this two-component channel spin formalism
is possible for more than two fragments in one or both channels.
For $n$ fragments in a channel, $n$-component channel spin
 wave functions can be introduced in the corresponding
 matrix elements and give again a possibility of using
 full antisymmetrization.
\bye